\documentclass[prd,aps,showpacs,tightenlines,nofootinbib,preprint]{revtex4}

\usepackage{epsf}
\usepackage{graphics}
\usepackage{amssymb}
\usepackage{amsfonts}
\usepackage[mathscr]{eucal}
\usepackage{amsmath}
\usepackage{enumerate}
\usepackage{hhline}
\usepackage{array}
\usepackage{tabularx}

\newcommand{\bear}{\begin{array}}  \newcommand{\eear}{\end{array}}
\newcommand{\bea}{\begin{eqnarray}}  \newcommand{\eea}{\end{eqnarray}}
\newcommand{\beq}{\begin{equation}}  \newcommand{\eeq}{\end{equation}}
\newcommand{\bef}{\begin{figure}}  \newcommand{\eef}{\end{figure}}
\newcommand{\bec}{\begin{center}}  \newcommand{\eec}{\end{center}}

\newcommand{\bib}{\bibitem}

\def\APJ#1#2#3{Astrophys. J. {\bf #1}, #2 (19#3)}
\def\APJJ#1#2#3{Astrophys. J. {\bf #1}, #2 (20#3)}

\def\ARAA#1#2#3{Annu. Rev. Astron. Astrophys. {\bf#1}, #2 (19#3)}

\def\MNRASS#1#2#3{Mon. Not. R. Astron. Soc. {\bf #1}, #2 (20#3)}

\def\NPB#1#2#3{Nucl. Phys. {\bf B#1}, #2 (19#3)}

\def\PLB#1#2#3{Phys. Lett. B {\bf #1}, #2 (19#3)}
\def\PLBB#1#2#3{Phys. Lett. B {\bf #1}, #2 (20#3)}
\def\PL#1#2#3{Phys. Lett. {\bf #1}, #2 (19#3)}

\def\PRD#1#2#3{Phys. Rev. D {\bf #1}, #2 (19#3)}
\def\PRDD#1#2#3{Phys. Rev. D {\bf #1}, #2 (20#3)}

\def\PRL#1#2#3{Phys. Rev. Lett. {\bf#1}, #2 (19#3)}

\def\PRT#1#2#3{Phys. Rep. {\bf#1}, #2 (19#3)}
\def\PRTT#1#2#3{Phys. Rep. {\bf#1}, #2 (20#3)}

\def\RPP#1#2#3{Rep. Prog. Phys. {\bf #1}, #2 (19#3)}

\def\RMPP#1#2#3{Rev. Mod. Phys. {\bf #1}, #2 (20#3)}

\def\Vec#1{\mbox{\boldmath $#1$}}
\def\Vecs#1{\mbox{\boldmath\tiny $#1$}}

\begin{document}

\title{Large-scale magnetic fields from inflation in dilaton electromagnetism}
\author{Kazuharu Bamba and J.\ Yokoyama}
\affiliation{Department of Earth and Space Science, Graduate School of
Science, Osaka University, Toyonaka 560-0043, Japan}

\date{\today}


\begin{abstract} 
The generation of 
large-scale magnetic fields is studied in dilaton 
electromagnetism in inflationary cosmology, taking into account the dilaton's
 evolution throughout inflation and reheating until it is stabilized
 with possible entropy production. 
It is shown that large-scale magnetic fields with 
observationally interesting strength at the present time could be generated 
if the conformal invariance of the Maxwell theory is broken through the 
coupling between the dilaton and electromagnetic fields
in such a way that the resultant quantum fluctuations in the magnetic field
have a nearly scale-invariant spectrum.  
If this condition is met, the amplitude of the generated magnetic field
 could be sufficiently large even in the case huge amount of entropy
is produced with the dilution factor $\sim 10^{24}$
as the dilaton decays.
\end{abstract}

\pacs{98.80.Cq, 98.62.En}
\hspace{13.0cm} OU-TAP 219


\maketitle


\section{Introduction}
It is well known that magnetic fields are present on various scales in the 
Universe, from planets, stars, galaxies, to clusters of galaxies (for recent 
detailed reviews see [1-3]). The origin of the cosmic magnetic fields, 
however, is not well understood yet. Since they have direct influence not 
only on various astrophysical situations but also on the evolution of the 
Universe, their origin is one of the most important problems in modern 
cosmology. 

In galaxies of all types, magnetic fields with the field strength 
$\sim 10^{-6}$G, ordered on $1-10$kpc scale, have been detected 
\cite{Widrow,Sofue}. There is some evidence that they exist in galaxies at 
cosmological distances \cite{Kronberg2}. Furthermore, in recent years 
magnetic fields in clusters of galaxies have been observed by means of 
the Faraday rotation measurements (RMs) of polarized electromagnetic radiation 
passing through an ionized medium \cite{Kim1}. Unfortunately, however, RMs 
inform us of only the value of the product of the field strength along the 
line of sight and the coherence scale, and so we cannot know the strength 
without assuming the value of the coherence scale and vice versa. 
In general, the strength and the scale are estimated on 
$10^{-7}-10^{-6}$G and 10kpc$-$1Mpc, respectively. It is very interesting 
and mysterious that magnetic fields in clusters of galaxies are as strong as 
galactic ones and that the coherence scale may be as large as $\sim$Mpc. 

Some elaborated magnetohydrodynamical (MHD) mechanisms have been proposed 
to amplify very weak seed magnetic fields into the $\sim 10^{-6}$G fields 
generally observed in galaxies. These mechanisms, known as\
\textit{galactic dynamo} \cite{EParker},\ are based on the conversion of 
the kinetic energy of the turbulent motion of conductive interstellar medium 
into magnetic energy. Galactic dynamo, however, is only an amplification 
mechanism, and so requires initial seed magnetic fields to feed on. Moreover, 
the effectiveness of the dynamo amplification mechanism in galaxies at high 
redshifts or clusters of galaxies is not well established. 

Scenarios for the origin of seed magnetic fields fall into two broad 
categories. One is astrophysical processes and the other is cosmological 
physical processes in the early Universe. The former, by and large, exploits 
the difference in mobility between electrons and ions. This difference can 
lead to electric currents and hence magnetic fields. The latter can also 
generate magnetic fields. Typically, magnetogenesis requires an 
out-of-thermal-equilibrium condition and a macroscopic parity violation. 
These conditions could have been naturally provided by the first-order 
cosmological electroweak phase transition (EWPT) \cite{Baym} or quark-hadron 
phase transition (QCDPT) \cite{Quashnock} (see more references in the review 
\cite{Grasso}). The bubbles of new phase were formed in the old one and 
strong, though small-scale, turbulent motion is excited in the plasma. 
In standard model, however, EWPT is the second order, so that such bubbles 
cannot be formed. 
Furthermore, it has recently been shown by Durrer and Caprini \cite{Durrer} 
that causally produced stochastic magnetic fields on large scales, {\it e.g.}, 
during EWPT or even later, are much stronger suppressed than usually assumed.  

If the scale of cluster magnetic fields is as large as $\sim$Mpc, it is likely 
that the origin of such a large-scale magnetic field is in physical processes 
in the early Universe rather than in astrophysical processes. From the two 
points, (1) There exists magnetic fields with the field strength 
$\sim10^{-6}$G even in the objects where the effectiveness of the dynamo 
amplification mechanism is not well established, and (2) There is the 
possibility that the scale of cluster magnetic fields is as big as 
$\sim$Mpc, it is conjectured that large-scale strong magnetic fields are 
produced in the early Universe and then are trapped in the plasma that 
collapsed to form galaxies and clusters of galaxies through adiabatic 
compression, or, in addition, secondary amplification mechanism such as 
galactic dynamo. 

Since the conductivity of the Universe through most of its history is large, 
the magnetic field $B$ evolves conserving magnetic flux as $B \propto a^{-2}$, 
where $a(t)$ is the scale factor. On the other hand, the average cosmic 
energy density $\bar{\rho}$ evolves as $\bar{\rho} \propto a^{-3}$ in the 
matter dominated epoch. Hence $B \propto {\bar{\rho}}^{2/3}$. The present 
ratio of interstellar medium density in galaxies ${\rho}_\mathrm{gal}$ to 
$\bar{\rho}$ and that of inter-cluster medium density in clusters of galaxies 
${\rho}_\mathrm{cg}$ are ${\rho}_\mathrm{gal}/\bar{\rho} \simeq 10^5-10^6$ and 
${\rho}_\mathrm{cg}/\bar{\rho} \simeq 10^2-10^3$, respectively. Consequently, 
from these relations, we see that the required strength of the cosmic magnetic 
field at the structure formation, adiabatically rescaled to the present time, 
is $10^{-10}-10^{-9} \mathrm{G}$ in order to explain the observed fields in 
galaxies $B_\mathrm{gal} \sim 10^{-6}$G and clusters of galaxies 
$B_\mathrm{cg} \sim 10^{-7}$G. 
On the other hand, in general, seed fields with the present strength 
$10^{-22}-10^{-16}$G is required for the galactic dynamo scenario. 

Although first-order cosmological phase transitions in the early Universe 
generate magnetic fields, the comoving coherence length of the magnetic fields 
cannot be larger than the Hubble horizon at the phase transition, which is 
much smaller than Mpc today. Though the coherence length may grow due to MHD 
effects, this happens at the expense of the magnetic field strength. 

The most natural mechanism overcoming the large-coherence-scale problem is 
\textit {inflation} in the early Universe (for a comprehensive introduction to 
inflation see Refs. \cite{Linde1,Kolb}). Turner and Widrow \cite{Turner} (TW) 
first indicated that large-scale magnetic fields could be generated in the 
inflationary stage. Inflation naturally produces effects on very large 
scales, larger than Hubble horizon, starting from microphysical processes 
operating on a causally connected volume. If electromagnetic quantum 
fluctuations are amplified during inflation, they could appear today as 
large-scale static magnetic fields. This idea is based on the assumption that 
a given mode is excited quantum mechanically while it is subhorizon sized and 
then as it crosses outside the horizon ``freezes in'' as a classical 
fluctuation. However, there is a serious obstacle on the way of this nice 
scenario as argued below. 

It is well known that quantum fluctuations of massless scalar and tensor 
fields are very much amplified in the inflationary stage and create 
considerable density inhomogeneities \cite{Guth} or relic gravitational waves 
\cite{Rubakov}. This is closely related to the fact that these fields are not 
conformally invariant even though they are massless. The amplification of the 
quantum fluctuations can be understood as particle production by an external 
gravitational field. Since the Friedmann-Robertson-Walker (FRW) metric 
usually considered is known to be conformally flat, the background 
gravitational field does not produce particles if the underlying theory is 
conformally invariant \cite{Parker}. This is the case for photons since the 
classical electrodynamics is conformally invariant in the limit of vanishing 
masses of fermions. Hence electromagnetic waves could not be generated in 
cosmological background. If the origin of large-scale magnetic fields in 
clusters of galaxies is electromagnetic quantum fluctuations generated and 
amplified in the inflationary stage, the conformal invariance must have been 
broken at that time. Several breaking mechanisms have been proposed, which 
are mainly classified into the following three types. 

(1) A non-minimal coupling of electromagnetic fields to gravity:\ 
TW introduced the gravitational couplings $RA_{\mu}A^{\mu}$, 
$R_{\mu\nu}A^{\mu}A^{\nu}$, $RF_{\mu\nu}F^{\mu\nu}/m^2$, etc, where $R$ is 
the curvature scalar, $A_{\mu}$ the electromagnetic potential, 
$F_{\mu\nu}$ the electromagnetic field-strength tensor, and $m$ a constant 
with dimension of mass. The $RA^2$ terms could generate large-scale magnetic 
fields with interesting strength, but they also break gauge invariance by 
giving the photon an effective mass. In contrast, the $RF^2$ terms are 
theoretically more plausible, but the strength of the resultant magnetic 
fields is very weak. 

(2) A coupling of a scalar field to electromagnetic fields:\ 
TW first indicated the coupling of an axion field, or that of a charged field 
which is not conformally coupled to gravity. After that many authors have 
studied more natural and effective couplings. 

Ratra \cite{Ratra} suggested the coupling of the inflation-driving scalar 
field (inflaton) $\phi$ in the form $e^{\omega\phi}F_{\mu\nu}F^{\mu\nu}$, 
and calculated the strength of large-scale magnetic fields in so-called 
a power-law inflation model induced by the exponential potential 
of the form $e^{\tilde{\omega}\phi}$, where $\omega$ and $\tilde{\omega}$ 
are constant parameters with dimension $(\mathrm{mass})^{-1}$. 
As a result, he found that present magnetic fields as large as 
$10^{-10}-10^{-9}$G could be generated. 
Recently Giovannini \cite{Giovannini} discussed the coupling of a massive 
scalar field $\varphi_\mathrm{m}$ other than the inflaton in the form 
$(\varphi_\mathrm{m}/M_{\mathrm{Pl}})^{\xi} F_{\mu\nu}F^{\mu\nu}$, where 
$\xi$ is a constant parameter and $M_{\mathrm{Pl}}$ is the Planck mass. 
According to Giovannini, large-scale magnetic fields with the strength larger 
than the dynamo requirement could be generated. 

Garretson, Field, and Carroll analyzed the amplification of electromagnetic 
fluctuations by their coupling to a pseudo Goldstone boson (PGB) 
${\varphi}_\mathrm{g}$ in the form 
${\varphi}_\mathrm{g} F_{\mu\nu}{\tilde{F}}^{\mu\nu}$, where 
${\tilde{F}}^{\mu\nu} \equiv 1/2 \hspace{1mm} 
{\varepsilon}^{\mu\nu\rho\sigma}F_{\rho\sigma}$ 
is the dual tensor of $F_{\mu\nu}$, and found that this coupling leads to 
exponential growth not for super-horizon modes but only for sub-horizon 
modes. Consequently, large-scale magnetic fields with interesting strength 
could not be generated \cite{Garretson}. 

Magnetic fields due to a charged scalar field were considered 
in a special model in \cite{Calzetta} (for more detailed review see 
\cite{Dolgov1}). The authors found that stochastic currents could be 
generated during inflation due to production of charged scalar particles by 
the inflaton, and in turn, magnetic fields. 
Moreover, Davis et al. argued that the backreaction of the scalar field gives 
the gauge field an effective mass thus breaking the conformal invariance 
\cite{Davis}. According to Davis et al., magnetic fields with the strength 
of order $10^{-24}$G on a scale of 100pc could be generated. 

(3) The conformal anomaly in the trace of the energy-momentum tensor induced 
by quantum corrections to Maxwell electrodynamics:\ It is known that the 
conformal anomaly, which is related to the triangle diagram connecting two 
photons to a graviton, breaks the conformal invariance by producing a 
nonvanishing trace of the energy-momentum tensor. 
Dolgov \cite{Dolgov2} pointed out that such an effect may lead to strong 
electromagnetic fields amplification during inflation. According to Dolgov, 
however, magnetic fields with interesting strength might not be generated in 
realistic case, {\it e.g.}, the model based on SU(5) gauge symmetry with 
three-fermion families. 

Incidentally, it has been indicated by Bertolami and Mota \cite{Bertolami} 
that the conformal invariance might be broken actually due to the possibility 
of spontaneous breaking of the Lorentz invariance in the context of string 
theories, and that generated large-scale magnetic fields could be strong 
enough to explain the observed fields through adiabatic compression. 

In the light of the above various suggestions, it seems at present that 
the most natural and effective way of breaking the conformal invariance is to 
introduce the coupling of a scalar field to electromagnetic fields.  In 
particular,  Ratra's 
suggestion is attractive who claimed present magnetic fields as large as 
$10^{-10}-10^{-9}$G could be generated in his model, which would not require 
any dynamo amplification to account for the observed fields in galaxies and 
clusters of galaxies. 

In the present paper, in addition to the inflaton field we assume the 
existence of the dilaton field and introduce the coupling of it to 
electromagnetic fields. Such 
coupling is reasonable in the light of indications in higher-dimensional 
theories, {\it e.g.}, string theories. Then we investigate the evolution of 
electromagnetic quantum fluctuations generated through the coupling, which 
breaks the conformal invariance of electrodynamics, and estimate the strength 
of large-scale magnetic fields at the present time. 
Particularly, we consider the following two cases. One is the case 
the dilaton freezes at the end of inflation in the same way as Ratra
\cite{Ratra} just for comparison, and the other is the more realistic
case that it still 
evolves after reheating and then decays into radiation with or without entropy 
production. 

Here we emphasize the following point. In Ratra's model, the inflaton and 
the dilaton are identified and power-law inflation is realized by introducing 
an exponential potential. There is no reason, however, why we should identify 
these fields. Furthermore, in the standard inflation models inflation is 
driven by the potential energy of the inflaton as it slowly rolls the 
potential hill. This slow roll over quasi-de Sitter stage is practically 
necessary to account for the nearly scale-invariant 
spectrum\footnote{The spectral index is estimated as $0.99 \pm 0.04$ by using 
the first year Wilkinson Microwave Anisotropy Probe (WMAP) data 
only \cite{Spergel}, where the errors are the 68\% confidence interval.} of 
the primordial curvature perturbation out of the quantum fluctuations of 
the inflaton. 

The reset of this paper is organized as follows. 
In Sec. II we describe our model action and derive the equations of motion 
from it. In Sec.\ III we investigate the evolution of electromagnetic 
fields, and then estimate the strength of the large-scale magnetic fields at 
the present time in Sec.\ IV, where
 we assume the dilaton freezes at the end of inflation. 
On the other hand, in Sec.\ V, we consider the case the dilaton 
still evolves after reheating and then decays into radiation 
with or without entropy production.\ Although we consider the evolution 
of electromagnetic fields in slow-roll exponential inflation models in 
Secs.\  II$-$V, for comparison we discuss it in power-law inflation models in 
Sec.\  VI keeping the recent WMAP data in mind.
Finally, Sec.\ VII is devoted to discussion and conclusion. 

We use units in which $k_\mathrm{B} = c = \hbar = 1$ and denote the 
gravitational constant $8 \pi G$ by ${\kappa}^2$ so that 
${\kappa}^2 \equiv 8\pi/{M_{\mathrm{Pl}}}^2$ where 
$M_{\mathrm{Pl}} = G^{-1/2} = 1.2 \times 10^{19}$GeV is the Planck mass.  
Moreover, in terms of electromagnetism we adopt Heaviside-Lorentz units.  
The suffixes `i', `1', `R', and `0' represent the quantities at the initial 
time $t_\mathrm{i}$, the time when a given mode first crosses the horizon 
during inflation $t_1$, the end of inflation (namely, the instantaneous 
reheating stage) $t_\mathrm{R}$, and the present time $t_0$, respectively.

\section{MODEL}

\subsection{Action}
We introduce two scalar fields, the inflaton field $\phi$, and the dilaton 
$\Phi$.  Moreover, we introduce the coupling of the dilaton to 
electromagnetic fields.  Our model action is given as follows.  
\begin{eqnarray}
 S &=& \int d^{4}x \sqrt{-g} \left[
                   \hspace{1mm}{\mathcal{L}}_{\mathrm{inflaton}}                             + {\mathcal{L}}_{\mathrm{dilaton}}  
              +  {\mathcal{L}}_{\mathrm{EM}} \hspace{1mm}
                 \right],  \label{eq:1}  \\[3mm]
 {\mathcal{L}}_{\mathrm{inflaton}} &=& -\frac{1}{2}g^{\mu\nu}{\partial}_{\mu}
              {\phi}{\partial}_{\nu}{\phi} 
               - U[\phi],  \label{eq:2} \\[3mm]
 {\mathcal{L}}_{\mathrm{dilaton}} &=& -\frac{1}{2}g^{\mu\nu}{\partial}_{\mu}
                                            {\Phi}{\partial}_{\nu}{\Phi} 
               - V[\Phi],  \label{eq:3} \\[3mm]
 {\mathcal{L}}_{\mathrm{EM}}  &=& 
                    -\frac{1}{4} f(\Phi) F_{\mu\nu}F^{\mu\nu}, 
           \label{eq:4} 
\end{eqnarray}   
\begin{eqnarray}
 f(\Phi) &=& \exp(\lambda \kappa \Phi),   
           \label{eq:5}  \\[3mm]
 V[\Phi] &=& \bar{V} \exp(-\tilde{\lambda} \kappa \Phi),
           \label{eq:6}
\end{eqnarray}   
where $g$ is the determinant of the metric tensor $g_{\mu\nu}$, 
$U[\phi]$ and $V[\Phi]$ are the inflaton and dilaton potentials, 
$\bar{V}$ is a constant, and $f$ is the coupling between the dilaton and 
electromagnetic fields with 
$\lambda$ and $\tilde{\lambda} \hspace{0.5mm} 
(\hspace{0.5mm} > 0 \hspace{0.5mm})$ being 
dimensionless constants.  
The form of the coupling between the dilaton and electromagnetic fields in 
Eq.\  (\ref{eq:5}) and that of the dilaton potential in Eq.\  (\ref{eq:6}) 
can be motivated by higher-dimensional theories reduced to four dimensions. 

We assume the spatially flat 
Friedmann-Robertson-Walker (FRW) space-time with the metric
\begin{eqnarray}
 {ds}^2 = g_{\mu\nu}dx^{\mu}dx^{\nu} =  -{dt}^2 + a^2(t)d{\Vec{x}}^2,
   \label{eq:7}
\end{eqnarray} 
where $a(t)$ is the scale factor.  
In terms of the U(1) gauge field $A_{\mu}$, the electromagnetic 
field-strength tensor is given by 
\begin{eqnarray}
 F_{\mu\nu} = {\partial}_{\mu}A_{\nu} - {\partial}_{\nu}A_{\mu}.
  \label{eq:8}
\end{eqnarray} 

Before going on, we state the framework we adopt in this paper.  

(1) During slow-roll inflation the cosmic energy density is dominated 
by $U[\phi]$ and the energy density of the dilaton is negligible.  

(2) The Universe is reheated immediately after inflation at $t=t_\mathrm{R}$.  
See {\it e.g.}\  \cite{KLS} for an efficient mechanism of reheating.  

(3) The conductivity of the Universe ${\sigma}_\mathrm{c}$ 
is negligibly small during inflation, because there are few charged particles 
at that time.  After reheating a number of charged particles are
 produced, so that the conductivity immediately jumps to a large value:\ 
${\sigma}_\mathrm{c}\gg H \hspace{1.5mm} 
(\hspace{0.5mm}t \geq t_\mathrm{R}\hspace{0.5mm})$,\ 
here $H$ is the Hubble parameter.  
This assumption is justified by a microphysical analysis \cite{Turner}.

(4) We consider both the case the dilaton is frozen at $t=t_\mathrm{R}$ 
as in Ratra's model \cite{Ratra} 
and that it continues to evolve after reheating until it decays with 
or without entropy production and gets stabilized at a potential minimum.  

(5) The value of the coupling $f$ between the dilaton and electromagnetic 
fields is set to unity when the dilaton gets stabilized so that the standard 
Maxwell theory is recovered:\ $f = 1$.  

\subsection{Equations of motion}
From the above action in Eq.\  (\ref{eq:1}) the equations of motion for the 
inflaton, the dilaton, and electromagnetic fields can be derived as follows.  
\begin{eqnarray}
&&-\frac{1}{\sqrt{-g}}{\partial}_{\mu}
             \left( \sqrt{-g}g^{\mu\nu}{\partial}_{\nu}
                               \phi \right)
               + \frac{dU[\phi]}{d\phi} = 0,  \label{eq:9} \\[3mm]
&&-\frac{1}{\sqrt{-g}}{\partial}_{\mu} \left(
                 \sqrt{-g}g^{\mu\nu}{\partial}_{\nu}
                               \Phi \right)
                    + \frac{dV[\Phi]}{d\Phi} 
                 = -\frac{1}{4}\frac{df(\Phi) }{d\Phi}  
                   F_{\mu\nu}F^{\mu\nu} \label{eq:10}, \\[3mm]
&&-\frac{1}{\sqrt{-g}}{\partial}_{\mu} \left(
                        \sqrt{-g} f(\Phi) F^{\mu\nu}                               \right) = 0.
\label{eq:11}  
\end{eqnarray} 

Since we are interested in the specific case where the background space-time 
is inflating, we assume that the spatial derivatives of 
$\phi$ and $\Phi$ are negligible compared to the other terms 
(if this is not the case at the beginning of inflation, any spatial 
inhomogeneities will quickly be inflated away and this assumption will
 quickly become very accurate).  Hence the equations of motion for the 
background homogeneous scalar fields read 
\begin{eqnarray}
&&\ddot{\phi} + 3H\dot{\phi} + \frac{dU[\phi]}{d\phi} = 0,
\label{eq:12} 
\end{eqnarray} 
\begin{eqnarray}
&&\ddot{\Phi} + 3H\dot{\Phi} + \frac{dV[\Phi]}{d\Phi} = 0,
\label{eq:13}  
\end{eqnarray} 
\hspace{0mm} together with the background Friedmann equation
\begin{eqnarray}
  H^2 &=&  \left( \frac{\dot{a}}{a} \right)^2 = \frac{{\kappa}^2}{3}
                 ({\rho}_{\phi} + {\rho}_{\Phi}),  
\label{eq:14} \\[3mm]
{\rho}_{\phi} &=& \frac{1}{2}{\dot{\phi}}^2 + U[\phi], 
\label{eq:15} \\[3mm]
{\rho}_{\Phi} &=& \frac{1}{2}{\dot{\Phi}}^2 + V[\Phi], 
\label{eq:16}
\end{eqnarray}
where a dot denotes a time derivative.  
Here ${\rho}_{\phi}$ and ${\rho}_{\Phi}$
 are the energy density of the inflaton 
and that of the dilaton.  
Since we have ${\rho}_{\phi} \gg {\rho}_{\Phi}$ by assumption, 
during inflation $H$ reads 
\begin{eqnarray}
 H^2 \approx \frac{{\kappa}^2}{3} U[\phi] \equiv {H_{\mathrm{inf}}}^2,
\label{eq:17}
\end{eqnarray} 
where $H_{\mathrm{inf}}$ is the Hubble constant in the inflationary stage.  

We consider the evolution of the gauge field in this background.  
Its equation of motion in the Coulomb gauge, 
$A_0(t,\Vec{x}) = 0$ and ${\partial}_jA^j(t,\Vec{x}) =0$, becomes 
\begin{eqnarray}
  \ddot{A_i}(t,\Vec{x})
 + \left( H + \frac{\dot{f}}{f} \right)
               \dot{A_i}(t,\Vec{x})
- \frac{1}{a^2}{\partial}_j{\partial}_jA_i(t,\Vec{x}) = 0
\label{eq:18}.  
\end{eqnarray}

\section{Evolution of magnetic fields}
In this section, we investigate the evolution of the vector 
potential and then consider that of the electric and magnetic fields.  

\subsection{Evolution of vector potential} 
To begin with, we shall quantize the vector potential $A_i(t,\Vec{x})$.  
It follows from the electromagnetic part of our model Lagrangian in 
Eq.\ (\ref{eq:4}) that the canonical momenta conjugate to the electromagnetic 
potential $A_{\mu}(t,\Vec{x})$ are given by 
\begin{eqnarray}
 {\pi}_0 = 0, \hspace{5mm} {\pi}_{i} = f(\Phi)a(t) \dot{A_i}(t,\Vec{x}).
\label{eq:19} 
\end{eqnarray}
We impose the canonical 
commutation relation between $A_i(t,\Vec{x})$ and ${\pi}_{j}(t,\Vec{x})$:\ 
\begin{eqnarray} 
  \left[ \hspace{0.5mm} A_i(t,\Vec{x}), {\pi}_{j}(t,\Vec{y}) 
  \hspace{0.5mm} \right] = i
 \int \frac{d^3 k}{{(2\pi)}^{3}}
             e^{i \Vecs{k} \cdot \left( \Vecs{x} - \Vecs{y} \right)}
        \left( {\delta}_{ij} - \frac{k_i k_j}{k^2 } \right),
\label{eq:20} 
\end{eqnarray}
where $\Vec{k}$ is comoving wave number, and $k$ denotes its amplitude 
$|\Vec{k}|$.  
From this relation we obtain the expression for $A_i(t,\Vec{x})$ as 
\begin{eqnarray} 
  A_i(t,\Vec{x}) = \int \frac{d^3 k}{{(2\pi)}^{3/2}}
  \left[ \hspace{0.5mm} \hat{b}(\Vec{k}) 
        A_i(t,\Vec{k})e^{i \Vecs{k} \cdot \Vecs{x} }
       + {\hat{b}}^{\dagger}(\Vec{k})
       {A_i}^*(t,\Vec{k})e^{-i \Vecs{k} \cdot \Vecs{x}} \hspace{0.5mm} \right],
\label{eq:21} 
\end{eqnarray}
where $\hat{b}(\Vec{k})$ and ${\hat{b}}^{\dagger}(\Vec{k})$ 
are the annihilation and creation operators which satisfy 
\begin{eqnarray} 
\left[ \hspace{0.5mm} \hat{b}(\Vec{k}), {\hat{b}}^{\dagger}({\Vec{k}}^{\prime}) \hspace{0.5mm} \right] = 
{\delta}^3 (\Vec{k}-{\Vec{k}}^{\prime}), \hspace{5mm}
\left[ \hspace{0.5mm} \hat{b}(\Vec{k}), \hat{b}({\Vec{k}}^{\prime})
\hspace{0.5mm} \right] = 
\left[ \hspace{0.5mm} 
{\hat{b}}^{\dagger}(\Vec{k}), {\hat{b}}^{\dagger}({\Vec{k}}^{\prime})
\hspace{0.5mm} \right] = 0.
\label{eq:22} 
\end{eqnarray}

From now on we choose the $x^1$ axis to lie along the spatial momentum 
direction \Vec{k} and denote the transverse directions $x^{I}$ with 
$I= 2,3$.  From Eq.\ (\ref{eq:18}) we find that the Fourier modes of the 
vector potential $A_i(t,k)$ satisfy the following equation:\ 
\begin{eqnarray} 
\ddot{A_I}(t,k) + \left( H_{\mathrm{inf}} + \frac{\dot{f}}{f} \right)
               \dot{A_I}(t,k) + \frac{k^2}{a^2} A_{I}(t,k) = 0, 
\label{eq:23}  
\end{eqnarray} 
and that the normalization condition for $A_i (t,k)$ reads
\begin{eqnarray} 
A_i(t,k){\dot{A}}_j^{*}(t,k) - {\dot{A}}_j(t,k){A_i}^{*}(t,k)
= \frac{i}{fa} \left( {\delta}_{ij} - \frac{k_i k_j}{k^2 } \right).
\label{eq:24} 
\end{eqnarray}

For convenience in finding the solutions of Eq.\ (\ref{eq:23}), we 
introduce the following approximate form as the expression of $f$.  
\begin{eqnarray}
f(\Phi) = f[\Phi(t)] = f[ \hspace{0.5mm} \Phi (a(t)) \hspace{0.5mm} ]
        \equiv \bar{f}a^{\beta-1},
\label{eq:25}
\end{eqnarray} 
where $\bar{f}$ is a constant and $\beta$ a parameter whose time-dependence 
is weak as will be seen in the next subsection.  
Using Eq.\ (\ref{eq:25}), we find 
\begin{eqnarray}
       H_{\mathrm{inf}} + \frac{\dot{f}}{f} = \beta H_{\mathrm{inf}}
\hspace{0.3mm}.  
\label{eq:26}
\end{eqnarray}
It follows from Eq.\ (\ref{eq:26}) that Eq.\ (\ref{eq:23}) is rewritten 
to the following form by replacing the independent variable $t$ to $\eta$.  
\begin{eqnarray}
 \frac{d^2 {A_I}(k,\eta)}{d{\eta}^2}  
        + \left( \frac{1-\beta}{ \eta } \right)
          \frac{d {A_I}(k,\eta)}{d \eta}
        + k^2 {A_I}(k,\eta) = 0, 
\label{eq:27}
\end{eqnarray}
where $\eta = \int dt/a(t)$ is conformal time.  
During inflation $\eta = -1/(aH_{\mathrm{inf}}$).  If we regard $\beta$ as 
a constant, the solution is given by 
\begin{eqnarray}
 A_I(k,\eta) = C_{I+}(k) (- H_{\mathrm{inf}} \eta)^{\beta/2}
H_{\beta/2}^{(1)} (-k \eta) + 
C_{I-}(k) (- H_{\mathrm{inf}} \eta)^{\beta/2}
H_{\beta/2}^{(2)} (-k \eta),
 \label{eq:28} 
\end{eqnarray} 
where $H_{\nu}^{(n)}$ is an $\nu$-th order Hankel function of type $n$ 
($n = 1,2$), and $C_{I+}(k)$ and $C_{I-}(k)$ are constants which satisfy
\begin{eqnarray}
|C_{I+}(k)|^2 - |C_{I-}(k)|^2 = \frac{\pi}{4H_{\mathrm{inf}}\bar{f}}
\hspace{0.5mm}.
\label{eq:29} 
\end{eqnarray} 
We shall choose
\begin{eqnarray}
C_{I+}(k) = \sqrt{ \frac{\pi}{4H_{\mathrm{inf}}\bar{f}} } \hspace{1mm}
   e^{i(\beta+1)\pi/4},  \hspace{5mm} 
C_{I-}(k) = 0,
\label{eq:30}
\end{eqnarray} 
so that the vacuum reduces to the one in Minkowski space-time at the 
short-wavelength limit:\ 
\begin{eqnarray}
\frac{k}{a H_{\mathrm{inf}}} = -k\eta 
      \to \infty  \hspace{1.5mm};\hspace{5mm} 
        A_I(k,\eta) \to
\frac{1}{\sqrt{2kf}} e^{-ik\eta}.
\label{eq:31} 
\end{eqnarray} 
We therefore obtain
\begin{eqnarray}
A_I(k,a) = \sqrt{\frac{\pi}{4H_{\mathrm{inf}}f(a)}}  a^{-1/2}
     H_{\beta/2}^{(1)} \left( \frac{k}{a H_{\mathrm{inf}}} \right) 
     e^{i(\beta+1)\pi/4}.   
\label{eq:32}
\end{eqnarray}

Being interested in large-scale magnetic fields, we investigate the 
behavior of this solution in the large-wavelength limit.  
Expanding the Hankel function in Eq.\ (\ref{eq:32}) and taking the first 
leading order in $k/(a H_{\mathrm{inf}})$, we obtain
\begin{eqnarray}
  A_I(k,a) &=& 2^{\beta/2}
                \sqrt{ \frac{1}{4 \pi H_{\mathrm{inf}}f(a)} } 
                \Gamma \left( \frac{\beta}{2} \right) a^{-1/2}
       \left( \frac{k}{aH_{\mathrm{inf}}}  \right)^{-\beta/2}      
       e^{i(\beta-1)\pi/4} \nonumber \\[3mm]
 &{\propto}& k^{-\beta/2} \hspace{1mm} a^0, \hspace{15mm}
\mathrm{for} \hspace{1.4mm} \beta >0, 
\label{eq:33} 
\end{eqnarray}
\hspace{0mm} and 
\begin{eqnarray}
 \hspace{3mm} A_I(k,a) &=& 2^{-\beta/2}
               \sqrt{ \frac{1}{4 \pi H_{\mathrm{inf}}f(a)} }  
               \Gamma \left( -\frac{\beta}{2} \right) a^{-1/2}
       \left( \frac{k}{aH_{\mathrm{inf}}}  \right)^{\beta/2}      
       e^{i(3-\beta)\pi/4}  \nonumber \\[3mm]
 \hspace{3mm} &{\propto}& k^{\beta/2}\hspace{1mm}a^{-\beta}, \hspace{15mm}
\mathrm{for} \hspace{1.4mm} \beta <0.
\label{eq:34}
\end{eqnarray}
The large-scale expansion bifurcates at $\beta = 0$.  This is the reason for 
the two different expressions, Eqs.\ (\ref{eq:33}) and (\ref{eq:34}).  
From Eqs.\ (\ref{eq:33}) and (\ref{eq:34}), we see that the large-scale 
vector potential is time-independent for $\beta >0$ and evolves like 
$a^{-\beta}$ for $\beta <0$.  Furthermore the mean square of the large-scale 
vector potential in the position space is 
$\sim k^3 |A_I(k,a)|^2\propto k^{-|\beta|+3}$ on a comoving scale 
$r = 2\pi/k$, so the root-mean-square (rms) has a scale-invariant spectrum 
when $|\beta| = 3$.  

\subsection{Parameter $\beta$} 
In the previous subsection, for convenience we have introduced 
the form (\ref{eq:25}) as the expression of $f$.  In practice, however, 
it is a function of $\Phi$ as seen in (\ref{eq:5}).  
We therefore investigate the expression of parameter $\beta$ by the comparison 
of Eq.\ (\ref{eq:5}) with Eq.\ (\ref{eq:25}).  

To begin with, we consider the evolution of the scale factor $a(t)$ and 
the dilaton $\Phi(t)$ in the inflationary stage.  The scale factor $a(t)$ 
is given by 
\begin{eqnarray}
     a(t) = a_1\exp[ \hspace{0.5mm} H_{\mathrm{inf}}(t-t_1)
                     \hspace{0.5mm}], 
\label{eq:35}
\end{eqnarray}
where $a_1$ is the scale factor at the time $t_1$ when a given comoving wavelength $2\pi/k$ of the vector potential first crosses outside the horizon during 
inflation, $k/(a_1 H_{\mathrm{inf}}) = 1$.  
In order to obtain the analytic solution of Eq.\ (\ref{eq:13}) 
we apply slow-roll approximation to the dilaton, that is, 
\begin{eqnarray}               
  \left| \frac{\ddot{\Phi}}{H_{\mathrm{inf}}\dot{\Phi}} \right| \ll 1,  
\label{eq:36}
\end{eqnarray}
and then Eq.\ (\ref{eq:13}) is reduced to 
\begin{eqnarray}   
 3H_{\mathrm{inf}} \dot{\Phi} +  \frac{dV[\Phi]}{d\Phi} = 0. 
\label{eq:37}
\end{eqnarray}
The solution of this equation is given by 
\begin{eqnarray}                             
  \Phi = \frac{1}{\tilde{\lambda}\kappa}
  \ln \left[ \hspace{0.5mm}  
     \frac{ (\tilde{\lambda}\kappa)^2\bar{V} }{3H_{\mathrm{inf}}}
      (t-t_\mathrm{R})  \hspace{0.5mm} + \exp( \tilde{\lambda} \kappa 
    {\Phi}_\mathrm{R})
      \right], 
\label{eq:38}  
\end{eqnarray}
where ${\Phi}_\mathrm{R} \hspace{0.5mm} 
( \hspace{0.5mm} \leq 0 \hspace{0.5mm})$ is the value 
of $\Phi$ at the end of inflation.  In the case the dilaton is frozen at 
$t=t_\mathrm{R}$, we choose ${\Phi}_\mathrm{R}=0$ so that $f=1$ at that time.  

Next, we investigate the evolution of $\dot{f}/f$.  
Using Eqs. (\ref{eq:5}), (\ref{eq:17}) and (\ref{eq:38}), we find 
\begin{eqnarray}
\frac{\dot{f}}{f} 
    \hspace{1mm}= \hspace{1mm} 
          \frac{\lambda \tilde{\lambda} {\kappa}^2 V[\Phi]}
                         {3H_{\mathrm{inf}}}
    \hspace{1mm} = \hspace{1mm}  
    \lambda \tilde{\lambda} H_{\mathrm{inf}} w,
\label{eq:39} 
\end{eqnarray} 
where $w$ is defined as
\begin{eqnarray}
 w \equiv \frac{ V[\Phi]}{{\rho}_{\phi} } \approx 
\frac{ V[\Phi]}{ U[\phi]}.
\label{eq:40}
\end{eqnarray}
Since we have ${\rho}_{\phi} \gg {\rho}_{\Phi}$ 
by assumption, $w \ll 1$.  
Comparing Eq.\ (\ref{eq:26}) with Eq.\ (\ref{eq:39}), we find 
\begin{eqnarray}
 \beta \approx 1 +  \lambda \tilde{\lambda} w = 1 + X \epsilon,
\label{eq:41}
\end{eqnarray}
where $X$ and $\epsilon$ are defined as 
\begin{eqnarray}
 X &\equiv& \frac{\lambda}{\tilde{\lambda}},  
\label{eq:42} \\[3mm]
 \epsilon &\equiv& {\tilde{\lambda}}^2 w,
\label{eq:43} 
\end{eqnarray}
respectively.  

Though $U[\phi]$ is approximately constant in slow-roll exponential inflation, 
$V[\Phi]$ changes gradually.  As a result it follows from 
Eqs.\ (\ref{eq:6}), (\ref{eq:38}), (\ref{eq:40}), and (\ref{eq:43}) 
that the ratio of $w$ at $t=t_\mathrm{R}$ to that at $t=t_\mathrm{i}$ is 
\begin{eqnarray}
 \frac{w(t_\mathrm{R})}{w(t_\mathrm{i})} &\approx&
 \frac{V[{\Phi}(t_\mathrm{R})]}{V[{\Phi}(t_\mathrm{i})]} \nonumber \\[3mm]
&=& 1 - 
\frac{ (\tilde{\lambda}\kappa)^2 V[{\Phi}_\mathrm{R}] }
 {3 {H_{\mathrm{inf}}}^2 }
 H_{\mathrm{inf}}(t_\mathrm{R}-t_\mathrm{i})  
\nonumber \\[3mm]
&=& 1 - \epsilon(t_\mathrm{R}) 
H_{\mathrm{inf}}(t_\mathrm{R}-t_\mathrm{i}),
\label{eq:44}
\end{eqnarray}
where $\epsilon(t_\mathrm{R}) = {\tilde{\lambda}}^2 w(t_\mathrm{R})$.  
Since $H_{\mathrm{inf}}(t_\mathrm{R}-t_\mathrm{i}) = 
N(t_\mathrm{i} \rightarrow t_\mathrm{R})$ is about 50, where 
$N(t_\mathrm{i} \rightarrow t_\mathrm{R})$ is the number of $e$-folds during 
the period from $t=t_\mathrm{i}$ to $t=t_{\mathrm{R}}$, the ratio of 
$w(t_\mathrm{R})$ to $w(t_\mathrm{i})$ is about half in the case 
$\epsilon(t_\mathrm{R}) \approx 1/100$.  If $\epsilon \ll 1$, the variation 
in $w$ in the inflationary stage is small, and then we can regard 
$\beta$ as approximately constant.  

The slow-roll condition to the dilaton, Eq.\ (\ref{eq:36}), 
is equivalent to the following relation.  
\begin{eqnarray}
\left| \frac{\ddot{\Phi}}{H_{\mathrm{inf}}\dot{\Phi}} \right| \ll 1  
\hspace{2mm} \Longleftrightarrow \hspace{2mm}
    \epsilon \ll 1.
\label{eq:45}
\end{eqnarray}
If we assume ${\tilde{\lambda}} \sim \mathcal{O}(1)$, 
the relation $\epsilon \ll 1$ is satisfied during inflation 
because of $w \ll 1$.  

\subsection{Evolution of electric and magnetic fields} 
We consider the evolution of electric and magnetic fields.  
The proper electric and magnetic fields are given by 
\begin{eqnarray}
&&{E_i}^{\mathrm{proper}}(t,\Vec{x})
    = a^{-1}E_i(t,\Vec{x}) = -a^{-1}\dot{A_i}(t,\Vec{x}), 
\label{eq:46} \\[3mm]
&&{B_i}^{\mathrm{proper}}(t,\Vec{x})
    = a^{-1}B_i(t,\Vec{x}) = a^{-2}{\epsilon}_{ijk}{\partial}_j A_k(t,\Vec{x}),
\label{eq:47}    
\end{eqnarray} 
where $E_i(t,\Vec{x})$ and $B_i(t,\Vec{x})$ are the comoving electric and 
magnetic fields, and ${\epsilon}_{ijk}$ is the totally antisymmetric tensor
(\hspace{0.5mm}${\epsilon}_{123}=1$\hspace{0.5mm}).  

From Eqs.\ (\ref{eq:32}), (\ref{eq:46}), and (\ref{eq:47}) we find the Fourier 
components of the comoving electric and magnetic fields in the inflationary 
stage:\ 
\begin{eqnarray}
     {E_I}(k,a)
 &=&   \sqrt{\frac{\pi}{4H_{\mathrm{inf}}f(a)}} \hspace{1.2mm} 
        \left( \frac{k}{a} \right) a^{-1/2} 
     H_{\beta/2 - 1}^{(1)} \left( \frac{k}{a H_{\mathrm{inf}}} \right) 
     e^{i(\beta+1)\pi/4},  
\label{eq:48}\\[3mm]
    {B_I}(k,a)
 &=& -i(-1)^I
      \sqrt{\frac{\pi}{4H_{\mathrm{inf}}f(a)}} \hspace{1.2mm} 
      \left( \frac{k}{a} \right) a^{-1/2}
     H_{\beta/2}^{(1)} \left( \frac{k}{a H_{\mathrm{inf}}} \right) 
     e^{i(\beta+1)\pi/4}.  
\label{eq:49}
\end{eqnarray} 

We consider the case the dilaton freezes at the end of inflation and 
after instantaneous reheating the conductivity of the Universe 
${\sigma}_\mathrm{c}$ jumps to a value much larger than the Hubble parameter 
at that time.  The evolutionary equation of the vector potential for an 
electrically conducting plasma is given by 
\begin{eqnarray}
 &&\ddot{A_i}(t,\Vec{x}) + 
\left( \frac{\dot{a}}{a} + {\sigma}_\mathrm{c} \right)
        \dot{A_i}(t,\Vec{x})
 - \frac{1}{a^2}{\partial}_j{\partial}_jA_i(t,\Vec{x}) = 0. 
\label{eq:50}
\end{eqnarray}
The joining conditions at the transition from the inflationary stage 
(INF) to the radiation-dominated one (RD) at $t=t_\mathrm{R}$ are \cite{Ratra} 
\begin{eqnarray}
    {E_i}^{(\mathrm{RD})}(t_\mathrm{R}, \Vec{x})  &=& 
               \exp(-{\sigma}_\mathrm{c}  t_{\mathrm{R}} )    
    {E_i}^{(\mathrm{INF})}(t_{\mathrm{R}}, \Vec{x}), 
\label{eq:51} \\[3mm]
    {B_i}^{(\mathrm{RD})}(t_{\mathrm{R}}, \Vec{x})  &=&    
    {B_i}^{(\mathrm{INF})}(t_{\mathrm{R}}, \Vec{x}). 
\label{eq:52}          
\end{eqnarray} 
From these joining conditions, for a large enough conductivity at the 
instantaneous reheating stage, we see that the electric fields accelerate 
charged particles and dissipate.  In fact, we solve Eq.\ (\ref{eq:50}) 
in the large-conductivity limit:\ ${\sigma}_\mathrm{c} \gg H$, 
so that electric fields vanish and the proper magnetic fields evolve in 
proportion to $a^{-2}(t)$ in the radiation-dominated stage 
and the subsequent matter-dominated stage 
(\hspace{0.5mm}$t \geq t_\mathrm{R}$\hspace{0.5mm}) \cite{Ratra}.  

It follows from Eq.\ (\ref{eq:49}) that the Fourier components of the proper 
magnetic fields in the inflationary stage are given by 
\begin{eqnarray}
|{B_I}^{\mathrm{proper}}(t,k)|^2 
&\equiv& a^{-2} \left| B_I(k,a) \right|^2  \nonumber \\[5mm]
&=& a^{-2} \left( \frac{\pi}{4H_{\mathrm{inf}}f(a)} \right)
     \left( \frac{k}{a} \right)^2  \left( \frac{1}{a} \right) 
     H_{\beta/2}^{(1)} \left( \frac{k}{a H_{\mathrm{inf}}} \right) 
     H_{\beta/2}^{(2)} \left( \frac{k}{a H_{\mathrm{inf}}} \right). 
\label{eq:53} 
\end{eqnarray} 
Since we are interested in the scales much larger than the Hubble radius in 
the inflationary stage, we expand Eq.\ (\ref{eq:53}) in the large-scale 
limit.  Taking account of the above fact that the proper magnetic fields 
evolve in proportion to $a^{-2}(t)$ 
(\hspace{0.5mm}$t \geq t_\mathrm{R}$\hspace{0.5mm}), 
we find that the Fourier components of the large-scale proper magnetic fields 
are expressed as 
\begin{eqnarray}
&&|{B_I}^{\mathrm{proper}}(t,k)|^2  =
\frac{2^{\beta-2}}{\pi} {\Gamma}^2 \left( \frac{\beta}{2} \right)
     f^{-1}(a_\mathrm{R}) 
\nonumber \\[3mm]
&& \hspace{3.5cm} \times
   \left( \frac{1}{a_\mathrm{R} H_{\mathrm{inf}}} \right)
 { \left( \frac{k}{a_{\mathrm{R}} H_{\mathrm{inf}}} \right) }^{-\beta}
          {  \left( \frac{k}{a} \right) }^2
          {  \left( \frac{1}{a} \right) }^2, \hspace{10mm}
\mathrm{for} \hspace{1.4mm} \beta > 0,
\label{eq:54}
\end{eqnarray} 
\hspace{0mm}and
\begin{eqnarray}
&&|{B_I}^{\mathrm{proper}}(t,k)|^2   =
    \frac{2^{-(\beta+2)}}{\pi} {\Gamma}^2  \left( -\frac{\beta}{2} \right)
    f^{-1}(a_\mathrm{R}) 
\nonumber \\[3mm]
&& \hspace{3.5cm} \times
   \left( \frac{1}{a_\mathrm{R} H_{\mathrm{inf}}} \right)
   { \left( \frac{k}{a_{\mathrm{R}} H_{\mathrm{inf}}} \right) }^{\beta}
          {  \left( \frac{k}{a} \right) }^2
          {  \left( \frac{1}{a} \right) }^2, \hspace{10mm}
\mathrm{for} \hspace{1.4mm} \beta < 0, 
\label{eq:55}
\end{eqnarray} 
where we have only recorded the first leading term.  

Finally, the energy density of the large-scale magnetic fields 
in Fourier space is given by 
\begin{eqnarray}
{\rho}_B(t,k) =  
          \left|{B_I}^{\mathrm{proper}}(t,k) \right|^2 f(a). 
\label{eq:56}
\end{eqnarray} 
Multiplying ${\rho}_B(t,k)$ by phase-space density:\ $4\pi k^3/(2\pi)^3$, 
we obtain the energy density of the large-scale magnetic fields 
in the position space 
\begin{eqnarray}
{\rho}_B(L,t) = 
 \frac{k^3}{2{\pi}^2}
 |{B_I}^{\mathrm{proper}}(t,k)|^2  \hspace{0.3mm} f(a), 
\label{eq:57}
\end{eqnarray} 
on a comoving scale $L=2\pi/k$.  Note that the expressions in 
Eqs.\ (\ref{eq:54}) and (\ref{eq:55}) are for one transverse component of the 
magnetic fields and the total magnetic field contribution is twice larger.
Using Eqs.\ (\ref{eq:54}), (\ref{eq:55}), and (\ref{eq:57}), 
we find that the energy density of the large-scale magnetic fields at 
the present time $t_0$ is given by 
\begin{eqnarray}
{\rho}_B(L,t_0) = 
  \frac{2^{|\beta|-3}}{{\pi}^3} {\Gamma}^2 \left( \frac{|\beta|}{2} \right)
    {H_{\mathrm{inf}}}^4 
   \left( \frac{a_{\mathrm{R}}}{a_0}  \right)^4 
  { \left( \frac{k}{a_{\mathrm{R}} H_{\mathrm{inf}}} \right) }^{-|\beta|+5}. 
\label{eq:58}
\end{eqnarray} 
Hence the large-scale magnetic fields have a scale-invariant spectrum when 
$|\beta| = 5$.  From now on we shall take the present scale factor $a_0 = 1$. 

\subsection{Consistency}
During inflation the energy density of electric and magnetic fields 
should be smaller than that of the dilaton so that the evolution of the 
dilaton is governed by its (classical) potential.  

We define the ratio of the energy density of the electric and magnetic fields 
to that of the dilaton as follows.  
\begin{eqnarray}
\Upsilon (L, t) &\equiv& \frac{1}{ {\rho}_{\Phi} }
\left[ \hspace{0.5mm} {\rho}_B (L, t) + {\rho}_E (L, t) 
       \hspace{0.5mm} \right], 
\label{eq:59}  \\[3mm]
{\rho}_E(L,t) &=& 
 \frac{k^3}{2{\pi}^2}
 |{E_I}^{\mathrm{proper}}(t,k)|^2  \hspace{0.3mm} f(a), 
\label{eq:60}
\end{eqnarray}
where ${\rho}_E(L,t)$ is the energy density of the electric fields 
on comoving scale $L=2\pi/k$.
During inflation $\Upsilon$ must be smaller than 1, which we call 
\textit{the consistency condition}.  

Using Eqs.\ (\ref{eq:17}), (\ref{eq:48}), (\ref{eq:49}), (\ref{eq:59}), and 
(\ref{eq:60}) we find 
\begin{eqnarray}
&& \Upsilon \approx \frac{1}{3w} \left( \frac{H_\mathrm{inf}}{M_{\mathrm{Pl}}}
                                   \right)^2
              \left( \frac{k}{aH_{\mathrm{inf}}} \right)^5 
\nonumber \\[2.5mm] 
&& \hspace{12mm} \times 
\left[  H_{\beta/2}^{(1)} \left( \frac{k}{a H_{\mathrm{inf}}} \right) 
              H_{\beta/2}^{(2)} \left( \frac{k}{a H_{\mathrm{inf}}} \right) 
            + H_{\beta/2-1}^{(1)} \left( \frac{k}{a H_{\mathrm{inf}}} \right) 
              H_{\beta/2-1}^{(2)} \left( \frac{k}{a H_{\mathrm{inf}}} \right) 
     \right],
\label{eq:61}
\end{eqnarray}
where the approximate equality follows from the ratio of the energy density of 
the dilaton to that of the inflaton in the inflationary stage:\ 
${\rho}_{\Phi}/{\rho}_{\phi} \approx w $, see Eqs.\ (\ref{eq:16}) 
and (\ref{eq:40}).  

Here we note that the upper limit on $H_{\mathrm{inf}}$ is determined by the 
observation of the anisotropy of cosmic microwave background (CMB) radiation.  
Using the WMAP data on temperature fluctuation \cite{Spergel}, which is 
consistent with the Cosmic Background Explorer (COBE) data, we can obtain 
a constraint on $H_{\mathrm{inf}}$ from tensor perturbation 
\cite{Rubakov,Abbott}, 
\begin{eqnarray}
\frac{H_{\mathrm{inf}}}{M_{\mathrm{Pl}}} \leq 2 \times 10^{-5}.
\label{eq:62} 
\end{eqnarray} 
From this relation we find that the upper limit on $H_{\mathrm{inf}}$ is 
$2.4 \times 10^{14} $GeV.  

When we consider a given scale $2\pi/k$, the value of 
$k/(aH_{\mathrm{inf}})$ decreases as the Universe evolves.  
 Evaluating Eq.\ (\ref{eq:61}) at the horizon crossing 
$k/(aH_{\mathrm{inf}}) =1$, we find $\Upsilon < 10^{-10}w^{-1}$, and so the 
consistency condition is satisfied.  
In the long wavelength regime, expanding Eq.  (\ref{eq:61}) 
and taking the first leading order in $ k/(aH_{\mathrm{inf}})$, we find 
\begin{eqnarray}
&& \hspace{5mm} \Upsilon \approx \frac{1}{3 {\pi}^2 w}  
  \left( \frac{H_\mathrm{inf}}{M_{\mathrm{Pl}}} \right)^2  
\left( \frac{aH_{\mathrm{inf}}}{k} \right)^{\beta-5} 
\nonumber \\[2.5mm] 
&& \hspace{9mm} \times
\left[ \hspace{1mm} 2^{\beta}{\Gamma}^2 \left( \frac{\beta}{2}  \right)
   +2^{\beta-2}{\Gamma}^2 \left( \frac{\beta}{2}-1 \right)
        \left( \frac{aH_{\mathrm{inf}}}{k} \right)^{-2}  \hspace{1mm} 
        \right], \hspace{7mm}
\mathrm{for} \hspace{1.4mm} \beta>2,    
\label{eq:63} \\[3mm]
&& \hspace{5mm} \Upsilon \approx \frac{1}{3 {\pi}^2 w}  
  \left( \frac{H_\mathrm{inf}}{M_{\mathrm{Pl}}} \right)^2  
\left( \frac{aH_{\mathrm{inf}}}{k} \right)^{\beta-5} 
\nonumber \\[2.5mm] 
&& \hspace{9mm} \times   
 \left[ \hspace{1mm} 2^{\beta}{\Gamma}^2  \left( \frac{\beta}{2}  \right)
   +2^{-(\beta-2)}{\Gamma}^2  \left( -\frac{\beta}{2}+1 \right)
   \left( \frac{aH_{\mathrm{inf}}}{k} \right)^{-2(\beta-1)} \hspace{1mm} 
        \right], \hspace{1.5mm}
\mathrm{for} \hspace{1.4mm} 0<\beta<2,
\label{eq:64} 
\end{eqnarray}
\hspace{0mm}and 
\begin{eqnarray}
&& \hspace{3.5mm} \Upsilon \approx \frac{1}{3 {\pi}^2 w}  
  \left( \frac{H_\mathrm{inf}}{M_{\mathrm{Pl}}} \right)^2 
\left( \frac{aH_{\mathrm{inf}}}{k} \right)^{-(\beta+5)}
\nonumber \\[2.5mm]
&& \hspace{13mm} \times       
\left[ \hspace{1mm} 2^{-\beta}{\Gamma}^2 \left( -\frac{\beta}{2} \right)
\hspace{0mm} 
         +2^{-(\beta-2)}{\Gamma}^2 \left( -\frac{\beta}{2}+1 \right)
        \left( \frac{aH_{\mathrm{inf}}}{k} \right)^{2}  \hspace{1mm} 
        \right], \hspace{7mm}
\mathrm{for} \hspace{1.4mm} \beta<0,    
\label{eq:65}
\end{eqnarray}   
where the first term in the square parentheses is the magnetic contribution and  the second term the electric one.  $\Upsilon$ is smaller on larger scales for 
$-3 < \beta < 5$.

\section{Estimation of present large-scale magnetic fields}
In this section, we estimate the present strength of the large-scale magnetic 
fields.  Since we consider the case the dilaton is frozen at 
$t=t_\mathrm{R}$, we choose ${\Phi}_\mathrm{R}=0$ 
so that $f=1$ at that time.  

\subsection{estimation of present large-scale magnetic fields}
To estimate the present large-scale magnetic fields from  Eq.\ (\ref{eq:58}) 
we use the following values \cite{Kolb}.  
\begin{eqnarray}
 H_0 &=& 70 \hspace{0.3mm}h_{70} \hspace{1mm} \mathrm{km} \hspace{1mm} 
                            {\mathrm{s}}^{-1} \hspace{1mm} 
                       {\mathrm{Mpc}}^{-1} 
\hspace{1mm} \approx  \hspace{1mm} 2.26 \hspace{0.3mm}h_{70} 
       \times 10^{-18} \hspace{1mm} {\mathrm{s}}^{-1},
\label{eq:66} \\[3mm]
 {\rho}_{\phi} &=& \frac{{\pi}^2}{30} g_{*} {T_\mathrm{R}}^4 
               \hspace{1.7mm} 
         \hspace{5mm}(g_{*} \approx 200),  
\label{eq:67} \\[3mm] 
 N &=& 45 + \ln \left( \frac{L}{\mathrm{[Mpc]}} \right) + 
    \ln \left\{ \frac{ \left[ 30/({\pi}^2 g_{*} )  \right]^{1/12} 
               {{\rho}_{\phi}}^{1/4} }
          {10^{38/3}\mathrm{[GeV]}}  \right\},
\label{eq:68} \\[3mm] 
\frac{a_0}{a_{\mathrm{R}}} &=& \left( \frac{ g_{*} }{3.91}
                               \right)^{1/3}
\frac{ T_{\mathrm{R}} }{T_{ \gamma 0} } 
\hspace{1mm} \approx \hspace{1mm}
\frac{ 3.7 T_{\mathrm{R}}}{2.35\times10^{-13}[\mathrm{GeV}]} 
   \hspace{5mm}( T_{ \gamma 0} \approx 2.73 \mathrm{K}), 
\label{eq:69}
\end{eqnarray}
where $H_0$ is the present Hubble parameter 
(throughout this paper we use $h_{70}=1.00$ \cite{HST}), 
$g_*$ is the total number of degrees of freedom for relativistic particles 
at the reheating epoch, $T_{\mathrm{R}}$ is reheating temperature, 
$N$ is the number of $e$-folds between the time $t_1$
and the end of inflation $t_{\mathrm{R}}$, and 
$T_{\gamma 0}$ is the present temperature of CMB radiation.  

Applying Eqs.\ (\ref{eq:66})$-$(\ref{eq:69}) and 
$k/(a_{\mathrm{R}} H_{\mathrm{inf}}) = \exp(-N)$ to Eq.\ (\ref{eq:58}), 
we find 
\begin{eqnarray}
  {\rho}_B (L, t_0) \propto |B(L, t_0)|^2 \propto  
  \left( \sqrt{H_{\mathrm{inf}} M_\mathrm{Pl}} L \right)^{|\beta|-5} 
                    \left(  T_{ \gamma 0}  
\sqrt{ \frac{H_{\mathrm{inf}}}{M_\mathrm{Pl}} }
                      \right)^4.
\label{eq:70}
\end{eqnarray} 
From this relation we see that stronger magnetic fields could be generated 
in the case of a large $H_{\mathrm{inf}}$ and a large $|\beta|$.  The 
reason is as follows.  From Eqs.\ (\ref{eq:38}) and (\ref{eq:43}), we find 
$\Phi(t_\mathrm{i}) = (\tilde{\lambda}\kappa)^{-1} 
\ln [\hspace{0.5mm} 1-\epsilon H_{\mathrm{inf}}(t_\mathrm{R}-t_\mathrm{i}) 
\hspace{0.5mm}]$.  Furthermore, from $\dot{f}/f = (\beta-1)H_{\mathrm{inf}}$, 
we find that the rate of the change of $f$ is larger in the case of 
a large $|\beta|$ and a large $H_{\mathrm{inf}}$.  Hence the conformal 
invariance of the Maxwell theory is broken to a larger extent in the case of 
a large $H_{\mathrm{inf}}$ and a large $|\beta|$.  Practically, for 
$\beta >0$, the initial amplitude of electromagnetic quantum fluctuation 
becomes larger.  On the other hand, for $\beta <0$, although the initial 
amplitude smaller, the rate of the amplification is very high.  We will 
consider the detailed values of the present magnetic fields in Sec.\ IV C.  

\subsection{Upper limits on cosmological magnetic fields}
Upper limits on cosmological magnetic fields come from the 
following three sources (see more detailed explanations in 
\cite{Widrow,Kolatt}).  

(1)\ CMB anisotropy measurements:\ 
Homogeneous magnetic fields during the time of decoupling whose scales are 
larger than the horizon at that time cause the Universe to expand at different 
rates in different directions.  Since anisotropic expansion of this type 
distorts CMB, measurements of CMB angular power spectrum impose limits on the 
cosmological magnetic fields.  Barrow, Ferreira, and Silk \cite{Barrow} 
carried out a statistical analysis based on the 4-year COBE data for angular 
anisotropy and derived the following limit on the primordial magnetic fields 
that are coherent on scale larger than the present horizon.  
\begin{eqnarray}
 B_{\mathrm{cosmic}}^{(0)}
       <  5 \times10^{-9} \hspace{1mm}\mathrm{G}.
\label{eq:71}
\end{eqnarray}
Incidentally, Caprini, Durrer, and Kahniashvili \cite{Caprini} have recently 
investigated the effect of gravity waves induced by a possible 
helicity-component of a primordial magnetic field on CMB temperature 
anisotropies and polarization.  
According to them, the effect could be sufficiently large to be observable 
if the spectrum of the primordial magnetic field is close to 
scale invariant and if its helical component is stronger than 
$\sim 10^{-10}$G.  
Moreover, it has also been argued that if tangled magnetic fields of 
$\gtrsim 3 \times 10^{-9}$G exist on cosmological scales, their imprint on 
CMB anisotropy and polarization maybe detectable \cite{Subramanian}.
 
(2)\ Big Bang Nucleosynthesis (BBN):\ 
Magnetic fields that existed during the BBN epoch would affect the expansion 
rate, reaction rates, and electron density.  Taking all these effects into 
account in calculation of the element abundances, and then comparing the 
results with the observed abundances, one can set limits on the strength of 
the magnetic fields.  The limits on homogeneous magnetic fields on the BBN 
horizon size $\sim 1.4 \times 10^{-4} {h_{70}}^{-1}\mathrm{Mpc}$ are less than 
$10^{-6}$G in terms of today's values \cite{Grasso2}.  

(3)\ Rotation Measure (RM) observations:\ 
RM data for high-redshift sources can be used to constrain the large-scale 
magnetic fields.  For example,\ Vall$\acute{\mathrm{e}}$e \cite{Vallee} 
tested for an RM dipole in a sample of 309 galaxies and quasars.  The 
galaxies in this sample extended to $z \simeq 3.6$ though most of the objects 
were at $z \lesssim 2$.  
Vall$\acute{\mathrm{e}}$e derived an upper limit of 
$6\times10^{-10}
{\left( {n_{e}}_0 /10^{-7}\mathrm{cm^{-3}} \right)}^{-1}\mathrm{G}$, where 
${n_{e}}_0$ is the present mean density of thermal electrons, 
on the strength of uniform component of a cosmological magnetic field.  
Note that the average baryon density is estimated as
${n_{b}}_0 = (2.7\pm0.1) \times 10^{-7} \mathrm{cm^{-3}}$
 \cite{Spergel} for comparison.

\subsection{Results}
We show the results of the present large-scale magnetic fields calculated 
from Eq.\ (\ref{eq:58}).  As described in Sec.  I, primordial magnetic fields 
with the present strength $10^{-10}-10^{-9}$G are required to explain the 
observed fields in galaxies and clusters of galaxies through adiabatic 
compression.  On the other hand, for galactic dynamo scenario, the fields 
with the present strength $10^{-22}-10^{-16}$G are required.  

Here we define $\Theta(L, t_\mathrm{R})$ as the ratio of the energy density of 
the large-scale electric fields to that of the magnetic counterpart at the 
end of inflation $t_\mathrm{R}$, 
\begin{eqnarray}
  \Theta(L, t_\mathrm{R}) \equiv 
    \frac{ \left| {E_I}^{\mathrm{proper}}(L, t_\mathrm{R}) \right|^2 }
        { \left| {B_I}^{\mathrm{proper}}(L, t_\mathrm{R}) \right|^2}. 
\label{eq:72}
\end{eqnarray}
From Eqs.\ (\ref{eq:63})$-$(\ref{eq:65}), (\ref{eq:68}) and (\ref{eq:72}), 
we find 
\begin{eqnarray}
&&\hspace{0mm} \Theta(L, t_\mathrm{R}) = 
        \frac{{\Gamma}^2(\beta/2-1)}{{\Gamma}^2(\beta/2)}
          K^{-2}(L),      
 \hspace{20mm} \mathrm{for} \hspace{1.4mm} \beta>2,  
\label{eq:73} \\ [9mm]   
&&\hspace{0mm}    \Theta(L, t_\mathrm{R}) =  
         \frac{{\Gamma}^2(-\beta/2+1)}
                 {{\Gamma}^2(\beta/2)}
       K^{-2(\beta-1)}(L),          
\hspace{9mm}  \mathrm{for} \hspace{1.4mm}  0<\beta<2,      
\label{eq:74} 
\end{eqnarray}
\hspace{0mm}and
\begin{eqnarray}
\hspace{2mm} \Theta(L, t_\mathrm{R}) =  
          \frac{{\Gamma}^2(-\beta/2+1)}
                 {{\Gamma}^2(-\beta/2)}
           K^{2}(L),
 \hspace{20mm}  \mathrm{for} \hspace{1.4mm} \beta < 0,
\label{eq:75} 
\end{eqnarray} 
\hspace{0mm}where
\hspace{0mm}
\begin{eqnarray}
\hspace{3mm}
K(L) \equiv 
2e^{45} \left\{ \frac{ \left[ 30/({\pi}^2 g_{*} )  \right]^{1/12} 
               {{\rho}_{\phi}}^{1/4} }
          {10^{38/3}\mathrm{[GeV]}}  \right\} 
       \left( \frac{L}{\mathrm{[Mpc]}} \right).
\label{eq:76} 
\end{eqnarray} 
While the energy density of magnetic fields is dominant for $\beta>1$, 
that of the electric fields is dominant for $\beta<1$ on large scales.  
Particularly, in the case $\beta<0$, the energy density of the large-scale 
electric fields is much larger than that of the magnetic counterpart.  
When strong enough magnetic field is generated, the accompanying electric 
field is so strong that the consistency condition is not satisfied.  
The reason is as follows.  As noted in Sec.\ III A, for $\beta>0$, 
the vector potential is time-independent in the large-scale limit in the 
inflationary stage, conversely, for $\beta<0$, it evolves like $a^{-\beta}$.  
Hence the electric fields are not generated in the former case, but 
generated in the latter case.  Consequently, stronger magnetic fields could 
be generated in the case $\beta>1$.  

Figures 1 and 2 depict the curves in the $H_{\mathrm{inf}}-\beta$ parameter 
space on which the present magnetic fields on 1Mpc scale with each strength 
could be generated.  The former is for $\beta > 0$ and the latter for 
$\beta < 0$.  In Fig.\ 1, the shaded area illustrates the excluded region 
from the observation of the anisotropy of CMB.  
This region is decided by Eq.\ (\ref{eq:71}).  From Eqs.\ (\ref{eq:63}) and 
(\ref{eq:64}) we find that the consistency condition is satisfied in all 
the area.  On the other hand, in Fig.\ 2, the shaded area corresponds to 
$\Upsilon \geq 1$ and illustrates the excluded region from the consistency 
condition decided by Eq.\ (\ref{eq:65}).  The excluded region from CMB is 
included in this region.  From these figures we can find the following 
results.  For $\beta>0$, the magnetic fields on 1Mpc scale with the strength 
larger than $10^{-10}$G could be generated in the case 
$H_{\mathrm{inf}} > 1.2 \times 10^{6} \mathrm{GeV}$.  
Furthermore, the magnetic fields strong enough for the galactic dynamo 
scenario ($10^{-22}-10^{-16}$G) could be generated in the wide region of the 
parameter space.  
On the other hand, for $\beta<0$, the maximum strength of the fields 
are $5 \times 10^{-19}$G at most.  
Parenthetically we note that in all figures we depict $10^{-22}$G contours 
with solid lines to emphasize the critical value for the galactic dynamo 
scenario.  
As we can see from Eq.\ (\ref{eq:70}), when $\beta > 5$, 
the larger-scale field is the stronger, so that its strength is constrained 
from the observation of the anisotropy of CMB, on the other hand, 
when $\beta < 5$, the stronger is the smaller-scale field, so that its 
strength is constrained from the predictions of light element abundances 
from BBN.  The former is more stringent than the latter, and so all results 
shown in Figs.\ 1 and 2 satisfy the limit imposed by the latter.  
Here we show a few characteristic examples.  
When $\beta = 1$, \textit{i.e.}, in the ordinary Maxwell theory, 
the present strength of the magnetic field on 1Mpc scale is 
$2.1 \times 10^{-58}$G.  This value is independent of $H_{\mathrm{inf}}$.  
Moreover, when $\beta = 5$ (the magnetic field has a scale-invariant 
spectrum), the maximum value of the field is $1.8 \times10^{-11}$G in the 
case $H_{\mathrm{inf}} = 2.4 \times 10^{14}\mathrm{GeV}$.  

Finally, we note the following point.  As discussed in Sec.\ III B, we have 
assumed $\epsilon \ll 1$ and regarded $\beta$ as approximately constant, and 
analytically investigated the evolution of the electric and magnetic 
fields.  This approximate analysis is proper.  In fact we have taken 
account of the time-dependence of $\beta$ in the case 
$w (t_\mathrm{R}) \ll 1$ and 
${\tilde{\lambda}} \sim \mathcal{O}(1)$:\ 
$\beta \sim 1 + \lambda w (t_\mathrm{R}) /\left( 1+w (t_\mathrm{R}) 
H_{\mathrm{inf}}(t-t_\mathrm{R}) \right)$, which is derived by using 
Eqs.\ (\ref{eq:6}), (\ref{eq:38}), (\ref{eq:40}), and (\ref{eq:41}), 
and numerically solved Eq.\ (\ref{eq:27}) in the inflationary stage.  
As a result we have confirmed that the numerical results almost agrees with 
the analytic ones.

\section{Dilaton decay after reheating}
So far we considered the case the dilaton field freezes at the end of 
inflation and so the value of the coupling $f$ between the dilaton and 
electromagnetic fields is set to unity.  In practice, however, it is expected 
that the dilaton continues its evolution along with the exponential potential 
even after reheating but is finally stabilized when it feels other 
contributions to its potential from, say, gaugino condensation that generates 
a potential minimum.  As it reaches there, the dilaton starts oscillation 
with mass $m$ and finally decays into radiation.  Then we consider its 
potential minimum is located at $\Phi=0$ and $f=1$ there, and so the field 
amplitude evolves from 
${\Phi}_{\mathrm{R}} (\hspace{0.5mm}<0 \hspace{0.5mm})$ 
to nearby zero along with the exponential potential.  

During the coherent oscillation the energy density of the dilaton 
${\rho}_{\Phi}$ evolves as $a^{-3}(t)$, so that it decreases more slowly 
than that of the radiation produced by the inflaton 
${{\rho}_\mathrm{r}}^{\mathrm{(inf)}}$.  
If the Universe is radiation-dominated until the dilaton decays, 
the entropy per comoving volume remains practically constant.  
If ${\rho}_{\Phi}$ becomes dominant over 
${{\rho}_\mathrm{r}}^{\mathrm{(inf)}}$, significant amount of 
entropy is produced, which dilutes the energy density of magnetic fields.  

\subsection{Dilaton decay}
We regard the time $t_{\mathrm{osc}} \simeq m^{-1}$ as the 
epoch when the coherent oscillations commence.  
When $t > t_\mathrm{R}$, the Universe is radiation-dominated, 
and so we shall set $a(t)=a_\mathrm{R}(t/t_{\mathrm{R}})^{1/2}$.  

Before the field oscillation regime 
(\hspace{0.5mm}$t_\mathrm{R}<t<t_\mathrm{osc}$\hspace{0.5mm}), 
the dilaton evolves with the exponential potential (6).  
It is known that, if the dilaton evolves with the exponential potential 
for a sufficiently long time, the dilaton enters a scaling regime as it 
evolves down its potential \cite{Barreiro}.  In this scaling regime the 
friction term from the expansion of the Universe in the equation of motion 
balances the potential force allowing it to enter this scaling era.  If the 
Universe is radiation-dominated in this regime, therefore, the evolution of 
the dilaton potential is the same as that of the background radiation 
produced by the inflaton, which is proportional to $t^{-2}$.  

Then we have numerically solved the equation of motion for the dilaton 
along with the exponential potential, and found that the dilaton evolves 
near to zero before it enters a scaling era because the field amplitude 
at the end of inflation $ |{\Phi}_\mathrm{R}|$ is constrained to be 
relatively small.  The reason is as follows.  
When \hspace{0mm}$t_\mathrm{R}<t<t_\mathrm{osc}$\hspace{0mm}, 
the energy density of magnetic fields is enhanced through the coupling with 
the evolving dilaton.  The energy density of magnetic fields on all scales 
should remain much smaller than that of the dilaton so that (4) should not 
affect the evolution of the dilaton.  As we will show in Eq.\ (\ref{eq:91}) 
in the next subsection, the energy density of magnetic fields is more 
enhanced in the case of a large value of 
$\tilde{\lambda} \kappa |{\Phi}_\mathrm{R}|$.  
Hence the upper limit on $ \tilde{\lambda} \kappa |{\Phi}_\mathrm{R}|$ 
is determined from this condition.  We have therefore numerically solved 
the equation of motion for the dilaton along with the exponential potential 
and investigated the region of $ \tilde{\lambda} \kappa |{\Phi}_\mathrm{R}|$ 
in which this condition is satisfied.  As a result we have found that 
the value of $|{\Phi}_\mathrm{R}|$ must be relatively small as 
$|{\Phi}_\mathrm{R}| \sim 1/\kappa$ in the case 
$ \tilde{\lambda} \sim \mathcal{O}(1) $.  From now on we will deal with 
$ \tilde{\lambda} \kappa |{\Phi}_\mathrm{R}|$ as a model parameter, and show 
the detailed region of this parameter in Figs.\ 3 and 4 in Sec.\ V C.  

Consequently, even if the energy density of radiation 
${\rho}_\mathrm{r}^{(\mathrm{inf})}$ is dominant over that of 
the dilaton ${\rho}_\mathrm{\Phi}$ at the instantaneous reheating stage, 
that is, 
$\bar{V}\exp (\hspace{0.3mm} -\tilde{\lambda} \kappa {\Phi}_\mathrm{R}
\hspace{0.3mm} )/{\rho}_{\phi} \approx w \ll 1$, 
${\rho}_\mathrm{r}^{(\mathrm{inf})}$ becomes 
comparable to ${\rho}_\mathrm{\Phi}$ around the field oscillation regime 
(\hspace{0.5mm}$ t \sim t_\mathrm{osc}$\hspace{0.5mm}).  
Since ${\rho}_\mathrm{\Phi}$ does not become dominant over 
${\rho}_\mathrm{r}^{(\mathrm{inf})}$ in 
\hspace{0mm}$t_\mathrm{R}<t<t_\mathrm{osc}$\hspace{0mm},\ however, 
it is the field oscillation regime 
that significant amount of entropy could be produced.  

We consider the evolution of ${\rho}_{\Phi}$ and the energy 
density of the radiation produced by the dilaton decay 
${{\rho}_\mathrm{r}}^{\mathrm{(dil)}}$ in the epoch of the coherent 
oscillations (\hspace{0.5mm}$t \ge t_{\mathrm{osc}}$\hspace{0.5mm}).  
The equations for them are given as follows \cite{Kolb}.  
\begin{eqnarray}
      \dot{{\rho}_{\Phi}} &=& 
        -\left( 3 \hspace{0.2mm} \frac{\dot{a}}{a} \hspace{0.2mm} 
+ {\Gamma}_{\Phi} \right) {\rho}_{\Phi}, 
\label{eq:77} \\[3.5mm] 
  \dot{{\rho}_{\mathrm{r}}}^{(\mathrm{dil})} &=& 
 -4 \hspace{0.2mm} \frac{\dot{a}}{a} \hspace{0.2mm} 
        {\rho}_{\mathrm{r}}^{(\mathrm{dil})}
       + {\Gamma}_{\Phi} {\rho}_{\Phi},
\label{eq:78}
\end{eqnarray} 
where ${\Gamma}_{\Phi}$ is the decay width of $\Phi$.  The solutions of 
these equations are given by 
\begin{eqnarray}
      {\rho}_{\Phi} &=& 
         {\rho}_{\Phi}(t_{\mathrm{osc}})
     \left[ \frac{a(t)}{a(t_{\mathrm{osc}})} \right]^{-3} 
 \exp \left[\hspace{0.5mm} - {\Gamma}_{\Phi} (t- t_{\mathrm{osc}}) 
        \hspace{0.5mm} \right],
\label{eq:79} \\[4.5mm]
   {{\rho}_{\mathrm{r}}}^{(\mathrm{dil})} &=&
  {\Gamma}_{\Phi}{\rho}_{\Phi}(t_{\mathrm{osc}})
 \left[ \frac{a(t)}{a(t_{\mathrm{osc}})} \right]^{-4}
\int_{t_{\mathrm{osc}}}^{t}
  \left[ \frac{a(\tau)}{a(t_{\mathrm{osc}})} \right]
        \exp \left[ \hspace{0.5mm}- {\Gamma}_{\Phi} (\tau- t_{\mathrm{osc}}) 
         \hspace{0.5mm}  \right] d\tau.
\label{eq:80}
\end{eqnarray}
Here we estimate 
${\rho}_{\Phi}(t=t_{\mathrm{osc}}) \approx \bar{V}$ because $\bar{V}$ 
is the value of the exponential potential in the form (\ref{eq:6}) at 
$\Phi = 0$, and we expect the contribution from stabilization mechanism 
has the same order of magnitude.  

\subsection{Entropy production}
If the Universe is radiation-dominated until 
the dilaton decays $t_{\Phi} \simeq {{\Gamma}_{\Phi}}^{-1}$, 
we find the following relation.  
\begin{eqnarray}
     {{\rho}_{\mathrm{r}}}^{(\mathrm{inf})} (t_{\Phi}) > 
      {\rho}_{\Phi}(t_{\Phi})
 &\Longleftrightarrow&  
     {\rho}_{\phi}  \left[ \frac{a(t_{\Phi})}{a_\mathrm{R}} \right]^{-4}
         > {\rho}_{\Phi}(t_{\mathrm{osc}})
     \left[ \frac{a(t_{\Phi})}{a(t_{\mathrm{osc}})} \right]^{-3}   
        \nonumber \\[4mm]
&\Longleftrightarrow&  
    \frac{{\rho}_{\phi}}{\bar{V}} > {{\Gamma}_{\Phi}}^{-1/2}
                         {t_\mathrm{R}}^{-2} m^{-3/2}  \nonumber \\[4mm]
&\Longleftrightarrow&   
       \frac{{\rho}_{\phi}}{\bar{V}}
 >  \left( \frac{M_\mathrm{Pl}}{m} \right)
                     \left( \frac{2H_{\mathrm{inf}}}{m} \right)^2, 
\label{eq:81}
\end{eqnarray}
where the last relation is obtained by using 
${\Gamma}_{\Phi} \simeq m(m/M_\mathrm{Pl})^2$ and 
$t_\mathrm{R} \approx 1/(2H_{\mathrm{inf}})$.  
In this case the entropy per comoving volume remains constant.  
Hence the necessary condition of entropy production is 
\begin{eqnarray}
  \frac{{\rho}_{\phi}}{\bar{V}}
                   < \left( \frac{M_\mathrm{Pl}}{m} \right)
                     \left( \frac{2H_{\mathrm{inf}}}{m} \right)^2.
\label{eq:82}
\end{eqnarray} 
Here we consider the case 
${{\rho}_\mathrm{r}}^{\mathrm{(inf)}}$ becomes equal to ${\rho}_{\Phi}$
at $t=t_\mathrm{c}$ in the epoch of the coherent oscillations 
$(\hspace{0.5mm}t_\mathrm{osc} < t_\mathrm{c} < t_{\Phi}\hspace{0.5mm})$:\ 
\begin{eqnarray}
   {{\rho}_\mathrm{r}}^{\mathrm{(inf)}} (t_\mathrm{c}) 
  = {\rho}_{\Phi}(t_\mathrm{c}) 
\Longleftrightarrow \hspace{2mm}
{\rho}_{\phi}  \left[ \frac{a(t_\mathrm{c})}{a_\mathrm{R}} \right]^{-4}
       =    {\rho}_{\Phi}(t_{\mathrm{osc}})
     \left[ \frac{a(t_\mathrm{c})}{a(t_\mathrm{osc})} \right]^{-3}.
\label{eq:83} 
\end{eqnarray}  
From this equation we find 
\begin{eqnarray}
    t_\mathrm{c} \approx
              \left( \frac{{\rho}_{\phi}}{ \bar{V}}
              \right)^2  {t_\mathrm{R}}^4 \hspace{0.2mm}
                         {t_{\mathrm{osc}}}^{-3} 
             \approx  \left( \frac{{\rho}_{\phi}}{ \bar{V} } \right)^2
          \left( \frac{1}{2H_{\mathrm{inf}}} \right)^4 m^3,
\label{eq:84} 
\end{eqnarray}  
where the first approximate equality follows from 
${\rho}_{\Phi}(t_{\mathrm{osc}}) \approx \bar{V}$, and the second one 
from $t_{\mathrm{osc}} \simeq m^{-1}$ and 
$t_\mathrm{R} \approx 1/(2H_{\mathrm{inf}})$.  
When $t \geq t_\mathrm{c}$,\ ${\rho}_{\Phi}$ is dominant over 
${{\rho}_\mathrm{r}}^{\mathrm{(inf)}}$, so that the Universe is 
matter-dominated.  Thus we shall set 
$a(t) = a_\mathrm{c} (t/t_\mathrm{c})^{2/3} = a_\mathrm{R} 
{t_\mathrm{R}}^{-1/2} {t_\mathrm{c}}^{-1/6} t^{2/3}$, where the second 
equality follows from the joining condition of $a(t)$ at $t=t_\mathrm{R}$.  

We investigate the entropy per comoving volume after the dilaton decay.  
In general, the entropy per comoving volume is represented as 
$S=a^3( \hspace{0.3mm} \rho + p \hspace{0.3mm}) / T$, where $\rho$, $p$ 
and $T$ are the equilibrium energy density, pressure, and temperature, 
respectively.  It is given by \cite{Kolb} 
\begin{eqnarray}
     S^{4/3} &=& {S_\mathrm{c}}^{4/3}  
+ \hspace{1mm}\frac{4}{3}{\rho}_{\Phi}(t_\mathrm{c})
               {a_\mathrm{c}}^{4} \left[ 
 \frac{2{\pi}^2 \langle g_{*}\rangle}{45}
      \right]^{1/3} {\Gamma}_{\Phi}
 \int_{t_\mathrm{c}}^{t} \left[ \frac{a(\tau)}{a_\mathrm{c}} \right]
  \exp \left[\hspace{0.5mm} - {\Gamma}_{\Phi} (\tau - t_{\mathrm{c}})
               \hspace{0.5mm}   \right] d\tau
\label{eq:85} \\[4mm]
&\approx&
 {S_\mathrm{c}}^{4/3} \left[\hspace{0.5mm} 
     1+{\Gamma}_{\Phi} {t_\mathrm{c}}^{-2/3}
         \int_{0}^{\infty}
             (u+t_{\mathrm{c}})^{2/3}
        \exp \left( - {\Gamma}_{\Phi} u  \right) du 
             \hspace{0.5mm} \right] \nonumber \\[4mm]
&\approx& {S_\mathrm{c}}^{4/3} \left[ \hspace{0.5mm}
  1+\Gamma \left( \frac{5}{3} \right) \left( t_{\mathrm{c}} {\Gamma}_{\Phi}
                \right)^{-2/3}  \hspace{0.5mm} \right],
\label{eq:86} 
\end{eqnarray}  
where $S_\mathrm{c}$ is the entropy per comoving volume 
at $t=t_\mathrm{c}$ and $\langle g_{*}\rangle$ is the appropriately-averaged 
value of $g_{*}$ over the decay interval.  In the second approximation 
we have introduced the variable $u \equiv \tau - t_{\mathrm{c}}$ and 
calculated in the limit $u \to \infty$.  Moreover we have used the relation 
${{\rho}_\mathrm{r}}^{\mathrm{(inf)}} (t_\mathrm{c}) 
= {\rho}_{\Phi}(t_\mathrm{c})$ and the following equation:\ 
\begin{eqnarray}
S = \left[\hspace{1mm}  \frac{4}{3} 
                \left( 
 \frac{2{\pi}^2 g_{*}}{45}
      \right)^{1/3} a^4 {\rho}_{\mathrm{r}} \hspace{1mm} \right]^{3/4},
\label{eq:87} 
\end{eqnarray}
which is the general relation between entropy per comoving volume $S$
 and the energy density of radiation ${\rho}_{\mathrm{r}}$.  It follows from 
${\Gamma}_{\Phi} \simeq m(m/M_\mathrm{Pl})^2$,\ 
Eqs.\ (\ref{eq:84}) and (\ref{eq:86}) that the ratio of the entropy 
per comoving volume after decay to that before decay is written as 
\begin{eqnarray}
\Delta S &\equiv& \frac{S}{S_\mathrm{c}} \nonumber \\[2mm]
         &\approx& \left\{ \hspace{1mm}
   1+ \Gamma \left( \frac{5}{3} \right)  \left[ 
 \left(  \frac{\bar{V}}{{\rho}_{\phi}}  \right)
 \left( \frac{2H_{\mathrm{inf}}}{m} \right)^2
 \left( \frac{M_\mathrm{Pl}}{m} \right) \right]^{4/3} \hspace{1mm}
\right\}^{3/4}  \nonumber \\[3mm]
&\approx&
 \left( \frac{\bar{V}}{{\rho}_{\phi}}  \right)
 \left( \frac{2H_{\mathrm{inf}}}{m} \right)^2
 \left( \frac{M_\mathrm{Pl}}{m} \right), 
\label{eq:88}
\end{eqnarray}
where the second approximate equality follows from 
Eq.\ (\ref{eq:82}).  

If the entropy production occurs, the Universe should expand 
larger enough to cancel out the effect of the produced entropy.  
From this point of view, taking account of 
${\rho}_{\mathrm{r}} \propto a^{-4}S^{4/3}$, we find 
\begin{eqnarray}
 \left( \Delta S  \right)^{4/3} = \left( \frac{\tilde{a_0}}{a_0} \right)^{4},
\label{eq:89}
\end{eqnarray}
where $\tilde{a_0}$ is the present scale factor in the case with the 
entropy production.   

Finally, we consider the effect of the dilaton decay on the energy density of 
the present large-scale magnetic fields.  We again assume that after 
instantaneous reheating the conductivity immediately jumps to a large value 
and so the amplitude of the vector potential is fixed.  
It follows from Eqs.\ (\ref{eq:5}), (\ref{eq:54})$-$(\ref{eq:57}), 
(\ref{eq:88}), and (\ref{eq:89}) that the ratio of 
the energy density of the present large-scale magnetic 
fields in the case the dilaton still evolves after reheating 
and then decays into radiation, 
$\tilde{{\rho}_B}(L,t_0)$, to that in the case the dilaton freezes at the 
end of inflation ${\rho}_B(L,t_0)$ is written as 
\begin{eqnarray}
 \frac{\tilde{{\rho}_B}(L,t_0)}{{\rho}_B(L,t_0)} 
&=& f^{-1}(t_\mathrm{R}) \left( \Delta S  \right)^{-4/3} 
\label{eq:90} \\[3mm]
&\approx&
  \exp \left( \hspace{0.5mm}  - \tilde{\lambda} \kappa {\Phi}_\mathrm{R}
          X  \hspace{0.5mm} \right) 
\left[ \hspace{1mm} 
\left( \frac{\bar{V}}{{\rho}_{\phi}}  \right)
 \left( \frac{2H_{\mathrm{inf}}}{m} \right)^2
 \left( \frac{M_\mathrm{Pl}}{m} \right)  \hspace{1mm}
\right]^{-4/3}
\label{eq:91} \\[3mm]
&\approx&
\exp \left[ \hspace{0.5mm}  - \tilde{\lambda} \kappa {\Phi}_\mathrm{R}
          \left( X+\frac{4}{3} \right)  \hspace{0.5mm} \right] 
\left[ \hspace{1mm} w
 \left( \frac{2H_{\mathrm{inf}}}{m} \right)^2
 \left( \frac{M_\mathrm{Pl}}{m} \right)  \hspace{1mm}
\right]^{-4/3},
\label{eq:92}
\end{eqnarray}
where the last approximate equality follows from 
$\bar{V}/{\rho}_{\phi} \approx w \exp (\hspace{0.3mm} 
\tilde{\lambda} \kappa {\Phi}_\mathrm{R}\hspace{0.3mm} )$, 
see Eqs.\ (\ref{eq:6}) and (\ref{eq:39}).  
In the right-hand-side of Eq.\ (\ref{eq:90}), the first factor 
is the contribution of the coupling with the dilaton, through 
which the energy density of the magnetic fields is enhanced 
because ${\Phi}_\mathrm{R}<0$. On the other hand, the second one 
is that of the produced entropy, which dilutes the 
energy density of the magnetic fields.  

\subsection{Effect of dilaton decay}
From Eqs.\ (\ref{eq:58}) and (\ref{eq:92}) we estimate the strength of 
the present large-scale magnetic fields in the case the dilaton still 
evolves after reheating and then decays into radiation.  
Figures 3 and 4 depict the magnetic field strength on 1Mpc scale at the 
present time $\tilde{B}(t_0)$.  The former is for the case 
$H_\mathrm{inf} = 10^{14} \mathrm{GeV}$ and $m=10^{13} \mathrm{GeV}$, 
while the latter corresponds to the case 
$H_\mathrm{inf} = 10^{10} \mathrm{GeV}$ 
and $m=10^{9} \mathrm{GeV}$.  In both these cases entropy is produced 
(\textit{i.e.}\ $\Delta S>1$), and then the relation (\ref{eq:82}) 
is satisfied.  The amount of the produced entropy in the former case is 
smaller than that in the latter.  As described in Sec.\ V A, the 
region of the parameter $ \tilde{\lambda} \kappa |{\Phi}_\mathrm{R}|$ is 
constrained from the condition that the energy density of magnetic fields on 
all scales should remain much smaller than that of the dilaton before the 
field oscillation regime 
(\hspace{0.5mm}$t_\mathrm{R}<t<t_\mathrm{osc}$\hspace{0.5mm}).  
This condition is satisfied in the region of 
$ \tilde{\lambda} \kappa |{\Phi}_\mathrm{R}| $ on each line 
in these figures.  
Moreover, Fig.\ 5 depicts the curves in the $H_\mathrm{inf}-m$ parameter space 
on which the present magnetic fields on 1Mpc scale with each strength could be 
generated for the case $\beta \approx 5.0 $.  
In Fig.\ 5, the shaded area illustrates the region with 
$m > 2H_\mathrm{inf}$, where $t_\mathrm{R} > t_\mathrm{osc}$ 
and our analysis does not apply.  
In all the cases in Fig.\ 5, entropy is 
also produced. In this figure, we have taken the maximum of 
$ \tilde{\lambda} \kappa |{\Phi}_\mathrm{R}|$ for each case. 
  In Figs.\ 3$-$5, we have taken $w =0.01$ and 
${\tilde{\lambda}} \sim \mathcal{O}(1)$.

The energy density of the magnetic fields is more enhanced 
for larger $ \tilde{\lambda} \kappa |{\Phi}_\mathrm{R}|$ 
and  larger $X$, while it is more diluted in the case more entropy 
is produced, see Eqs.\ (\ref{eq:90})$-$(\ref{eq:92}).  
In fact, if $\beta \approx 1+ X \epsilon \approx 5.0$, that is, the spectrum 
of the magnetic fields is close to the scale-invariant one, and 
produced entropy is relatively small, {\it e.g.}, in the case 
$H_\mathrm{inf} = 10^{14} \mathrm{GeV}$ and $m = 10^{13} \mathrm{GeV}$, 
in which $\Delta S = 4.6 \times 10^6$, the generated
magnetic fields on 1Mpc scale
could be
stronger than $10^{-10}$G at the present time.
Furthermore, in the case $\beta \approx 5.0$, even if the  entropy
production factor is as large as $\Delta S \sim 10^{24}$, a sufficient
magnitude of magnetic fields for the galactic dynamo scenario, $B
\gtrsim 10^{-22}$G, could be generated.
More specifically, the amplitude of the generated magnetic field ranges
from $10^{-22}$G to $10^{-16}$G, namely the span favored by the dynamo
scenario, if the dilaton mass lies in the range  
$9.3 \times 10^{6} \mathrm{GeV} \leq m \leq 9.3 \times 10^{9} \mathrm{GeV}$   
in the case $H_\mathrm{inf} = 10^{14} \mathrm{GeV}$ and $\beta =5$ 
with the entropy 
production factor being $\Delta S = 5.8\times10^{15} - 5.8\times10^{24}$. 
In the case
$H_\mathrm{inf} = 10^{10} \mathrm{GeV}$ and $\beta=5$,
 the appropriate amplitude of
magnetic field results for
$1.7 \times 10^{4} \mathrm{GeV} \leq m \leq 1.7 \times 10^{7} \mathrm{GeV}$ 
in which $\Delta S$ ranges $9.2 \times 10^{15}- 9.2 \times 10^{24}$.

Moreover, as shown in Figs.\ 3 and 4, even if $\beta < 5$, 
that is, the spectrum of generated magnetic field is blue, sufficiently large 
seed field for the galactic dynamo scenario could also be 
generated.  

Finally, in the case $X$ is so small ({\it e.g.}, $X \lesssim 10$)
that $\beta$ is about unity, the spectrum is too blue for the
large-scale magnetic field to be strong enough for the dynamo scenario
to work.  This result remain unchanged even when
$ \tilde{\lambda} \kappa |{\Phi}_\mathrm{R}|$ is 
large and the field amplitude is enhanced with the relation (81) being
 satisfied, namely, without any entropy production.  
Thus the essential requirement is that the spectrum should not be too blue.

\section{case of power-law inflation}
So far we considered the evolution of electromagnetic fields in slow-roll 
exponential inflation models in the light of the fact that the recent data 
of WMAP favors models with small slow-roll parameters.  In this section, 
for comparison with \cite{Ratra}, we discuss it in power-law inflation models 
with the following exponential inflaton potential.  
\begin{eqnarray}
  U[\phi] = \bar{U} \exp(-\zeta \kappa \phi), 
\label{eq:93}
\end{eqnarray}
where $\bar{U}$ is a constant and $\zeta$ is a dimensionless constant.  
In this case the spectral index of curvature perturbation $n_s$ 
is given by 
\begin{eqnarray}
   n_s - 1 = -6 {\epsilon}_U + 2 {\eta}_U = - {\zeta}^2, 
\label{eq:94} 
\end{eqnarray}
\begin{eqnarray}
  {\epsilon}_U &\equiv& \frac{1}{2{\kappa}^2}  
      \left( \frac{U^{\prime}}{U} \right)^2, 
\label{eq:95} \\[3mm]
  {\eta}_U &\equiv& \frac{1}{{\kappa}^2}  
      \left( \frac{U^{\prime\prime}}{U} \right). 
\label{eq:96} 
\end{eqnarray}
According to the WMAP data \cite{Peiris}, $n_s \geq 0.93$, and hence 
$\zeta \leq 0.26$.  In this case the scale factor in the inflationary 
stage is given by $a(t) \propto t^p$, where $p=2/{\zeta}^2 \geq 29$.  

If power-law inflation lasts for a sufficiently long time, the dilaton 
will settle to the scaling solution \cite{Barreiro} where 
$U^{\prime\prime} \approx {H_\mathrm{inf}}^2$ with $H_\mathrm{inf} = p/t$.  
Hence the solution of the dilaton in this regime is given by 
\begin{eqnarray}
   \Phi = \frac{2}{\tilde{\lambda}\kappa}
  \ln \left( \hspace{0.5mm}  
     \frac{ \sqrt{\bar{V}} \tilde{\lambda}\kappa t }{p}
      \right).  
\label{eq:97}
\end{eqnarray}
From Eqs.\  (\ref{eq:5}) and (\ref{eq:97}) we find 
\begin{eqnarray}
   f(\Phi) =  \left( \frac{ \sqrt{\bar{V}} \tilde{\lambda}\kappa t }{p}
              \right)^{ 2\lambda/\tilde{\lambda} }
           \propto \hspace{1mm} a^{2\lambda/(\tilde{\lambda}p)}. 
\label{eq:98}
\end{eqnarray}
Comparing Eq.\ (\ref{eq:25}) with Eq.\ (\ref{eq:98}), we find 
\begin{eqnarray}
    \beta =   \frac{2\lambda}{\tilde{\lambda}p} +1. 
\label{eq:99}
\end{eqnarray}
On the other hand, in the inflationary stage the Fourier modes of the vector 
potential satisfy the following equation:\ 
\begin{eqnarray}
 \frac{d^2 {A_I}(k,\eta)}{d{\eta}^2}  
        +  \frac{2\lambda}{\tilde{\lambda}} 
        \left( \frac{-1}{p-1} \frac{1}{\eta} \right)
          \frac{d {A_I}(k,\eta)}{d \eta}
        + k^2 {A_I}(k,\eta) = 0, 
\label{eq:100}
\end{eqnarray}
where $\eta = \int dt/a(t)$ is conformal time.  
Comparing Eq.\ (\ref{eq:27}) with Eq.\ (\ref{eq:100}), we find 
\begin{eqnarray}
    \beta = \frac{2\lambda}{\tilde{\lambda} (p-1)} + 1. 
\label{eq:101}
\end{eqnarray}
Here the solution of Eq.\ (\ref{eq:100}) is given by the same form as 
Eq.\ (\ref{eq:28}).  Since $p \gg 1$ as noted above, we can approximately 
identify Eq.\ (\ref{eq:99}) with Eq.\ (\ref{eq:101}).  Thus under this 
approximation we can develop the argument in the same way as in 
Secs.\ III$-$V.  

As noted in Secs.  IV and V, where we considered the case of exponential 
inflation, if $\beta \approx 5$, that is, the spectrum 
of the magnetic fields is close to the scale-invariant one, magnetic fields 
on 1Mpc scale with the strength $10^{-10}-10^{-9}$G at the present time could 
be generated.  In the case $w = 0.01$ and 
${\tilde{\lambda}} \sim \mathcal{O}(1)$, if $\beta \approx 5$, 
$X = \lambda/\tilde{\lambda} \approx 400$.  On the other hand, 
it follows from Eqs.\ (\ref{eq:99}) and (\ref{eq:101}) that, 
in the case of power-law inflation, 
if $\beta \approx 5$, $X \approx 2p \geq 58$.  
Consequently, even in the case of power-law inflation with the maximal 
breaking of the scale invariance of primordial fluctuations within 
observational limit \cite{Peiris}, 
$X$ should be much larger than unity in order that the amplitude of the 
generated
magnetic fields could take $10^{-10}-10^{-9}$G on 1Mpc 
scale at the present time.
  That is, the value of $n_s$\ obtained by the WMAP data 
is so close to unity that the power-law inflation model observationally 
permitted is not practically very much different from slow-roll exponential 
inflation.  

\section{Conclusion}
In the present paper we have studied the generation of large-scale magnetic 
fields in inflationary cosmology, breaking the conformal invariance of the 
electromagnetic field by introducing a coupling with the dilaton field.  
First we  considered the case the dilaton freezes at the end of
inflation automatically as assumed by Ratra \cite{Ratra}, to see how the
recent detailed observation of the primordial spectrum of density
fluctuations in terms of CMB anisotropy \cite{Spergel}, which favors
slow-rollover inflation, affects Ratra's previous analysis.  As a result
we have found the resultant magnetic field could be as large as
$10^{-10}-10^{-9}$G on 1Mpc scale at present for $H_\mathrm{inf}\gtrsim 
10^{6} \mathrm{GeV}$ provided that the model parameters are so chosen
that the spectrum of the magnetic field nearly scale-invariant or even
red. 

Next we considered a more realistic case that the dilaton continues its
evolution with 
the exponential potential after inflation until it is stabilized after
oscillating around its potential minimum.  It has two distinct effects
on the final amplitude of the magnetic field.  That is, the energy
density of the magnetic field is enhanced as the dilaton evolves due to
the exponential coupling, while it could produce huge amount of entropy
as it decays at a late time with the gravitational interaction.  We have
parameterized the evolution of the dilaton in terms of its mass, $m$,
around the potential minimum and its amplitude at the end of inflation, 
$\Phi_{\rm R}$, and adopted a view that it starts oscillation at
$t\simeq m^{-1}$, since the detailed shape of the dilaton potential
around the minimum is not known due to the fact that the stabilization
mechanism of the dilaton is not fully established yet, although there
are a number of proposals.  As a result we have found that the magnetic
field could be as large as $10^{-10}$G even with the entropy increase
factor $\Delta S \sim 10^6$ provided that the scale of inflation is
maximal and the spectrum is close to scale invariant.  Furthermore the
seed field for the dynamo mechanism could be accounted for even when
$\Delta S$ is as large as $10^{24}$ if model parameters are chosen
appropriately to realize nearly scale-invariant spectrum.

Thus the possible dilution due to huge entropy production from decaying
dilaton is not the primary obstacle to account for the large-scale
magnetic field in terms of quantum fluctuations generated during
inflation in this dilaton electromagnetism.  
The more serious requirement is that the model parameters should be so
chosen that the spectrum of generated magnetic field should not be too
blue but close to the scale-invariant or the red one, which is realized
only if a huge hierarchy exists between $\lambda$ and $\tilde\lambda$,
namely, $X$ should be extremely larger than unity.  This may make 
it difficult to motivate this type of model in realistic high energy 
theories.  This feature is independent of whether one considers
slow-roll exponential inflation or power-law inflation with an
exponential inflaton potential as adopted by Ratra \cite{Ratra}, because
the latter model is hardly distinguishable from the former under the
constraint imposed by WMAP data as far as the evolution of the dilaton
is concerned.

\section*{Acknowledgements}
This work was partially supported by the JSPS Grant-in-Aid for 
Scientific Research No.13640285(JY).
 


\newpage

\begin{figure}[tbp]
\begin{center}
   \includegraphics{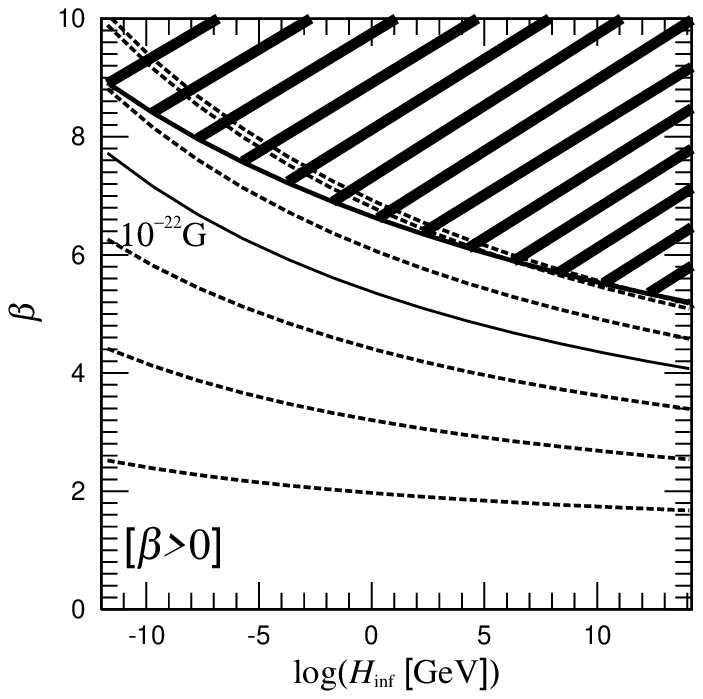}
\caption{The curves (dotted lines and a solid line) in the 
$H_\mathrm{inf}-\beta$ parameter space on which the present magnetic fields on 1Mpc scale with each strength could be generated 
(\hspace{0.5mm}$\beta>0$\hspace{0.5mm}). 
$B(t_0)=10^{-9} \mathrm{G}$, $10^{-10} \mathrm{G}$, $10^{-16} \mathrm{G}$, $10^{-22} \mathrm{G}$ (solid line), $10^{-30} \mathrm{G}$, $10^{-40} \mathrm{G}$, 
and $10^{-50} \mathrm{G}$ are shown (top down). The shaded area illustrates 
the excluded region from the observation of the anisotropy of CMB, 
Eq.\ (\ref{eq:71}).
}


\vspace{15mm}

   \includegraphics{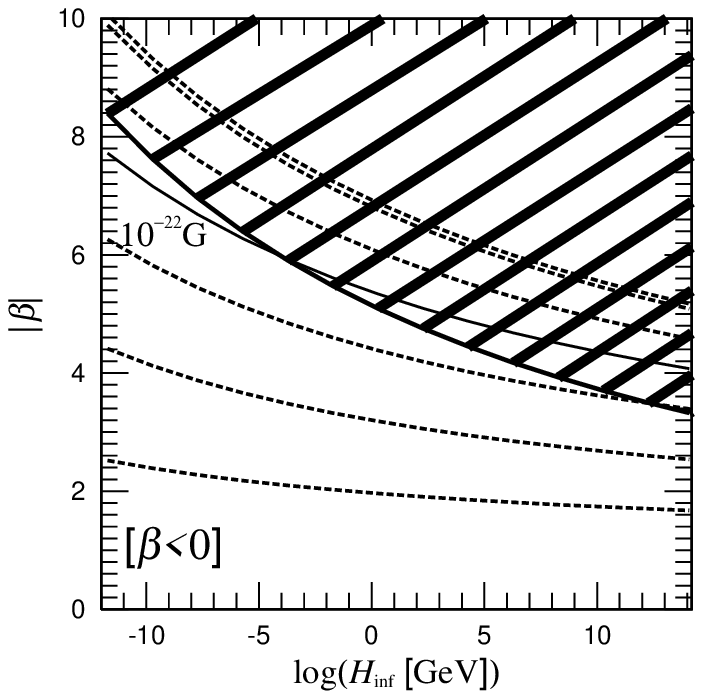}
\caption{The curves (dotted lines and a solid line) in the 
$H_\mathrm{inf}-\beta$ parameter space on which the present magnetic fields on 
1Mpc scale with each strength could be generated 
(\hspace{0.5mm}$\beta<0$\hspace{0.5mm}). 
$B(t_0)=10^{-9} \mathrm{G}$, $10^{-10} \mathrm{G}$, 
$10^{-16} \mathrm{G}$, $10^{-22} \mathrm{G}$ (solid line), 
$10^{-30} \mathrm{G}$, $10^{-40} \mathrm{G}$, and $10^{-50} \mathrm{G}$ are 
shown (top down). The shaded area corresponds to $\Upsilon \geq 1$ and 
illustrates the excluded region from the consistency condition decided by 
Eq.\ (\ref{eq:65}). This area includes the excluded region from the 
observation of the anisotropy of CMB. }
\end{center}
\end{figure}

\newpage

\begin{figure}[tbp]
\begin{center}
   \includegraphics{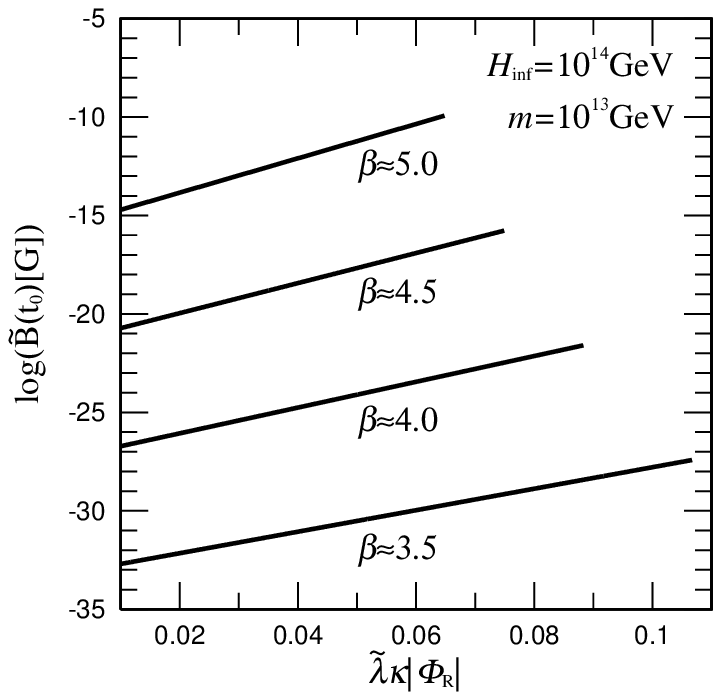}
\caption{The magnetic field strength on 1Mpc scale at the present time 
$\tilde{B}(t_0)$ in the case with entropy production. 
The lines are for the case $H_\mathrm{inf} = 10^{14} \mathrm{GeV}$ and 
$m=10^{13} \mathrm{GeV}$. 
$\beta \approx 1 +  \lambda \tilde{\lambda} w \approx 5.0$, 
$\beta \approx 4.5$,    
$\beta \approx 4.0$, and 
$\beta \approx 3.5$ are shown 
(top down). 
Here we have taken 
$w =0.01$ and ${\tilde{\lambda}} \sim \mathcal{O}(1)$. 
}


\vspace{15mm}

   \includegraphics{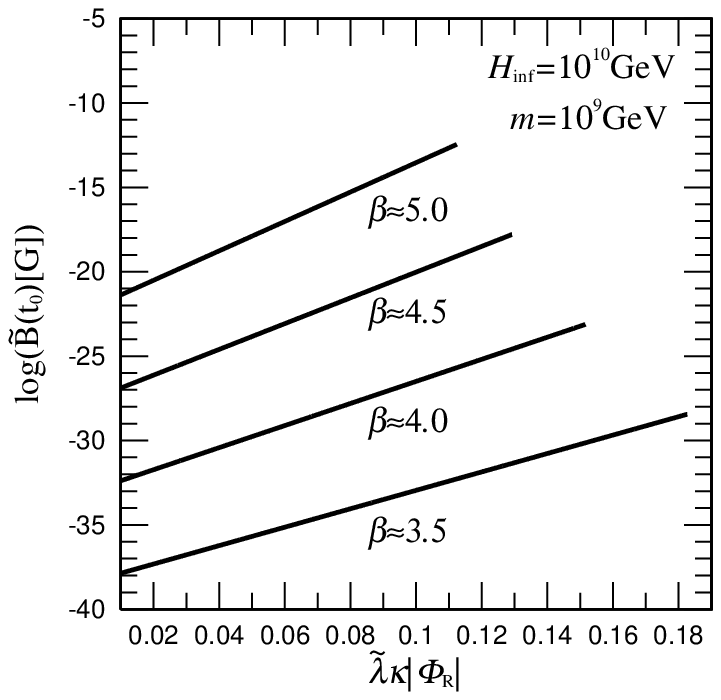}
\caption{The magnetic field strength on 1Mpc scale at the present time 
$\tilde{B}(t_0)$ in the case with entropy production. 
The lines are for the case $H_\mathrm{inf} = 10^{10} \mathrm{GeV}$ and 
$m=10^{9} \mathrm{GeV}$. 
$\beta \approx 1 +  \lambda \tilde{\lambda} w \approx 5.0$, 
$\beta \approx 4.5$,    
$\beta \approx 4.0$, and 
$\beta \approx 3.5$ are shown 
(top down). 
Here we have taken 
$w =0.01$ and ${\tilde{\lambda}} \sim \mathcal{O}(1)$. 
}
\end{center}
\end{figure}

\newpage

\begin{figure}[tbp]
\begin{center}
   \includegraphics{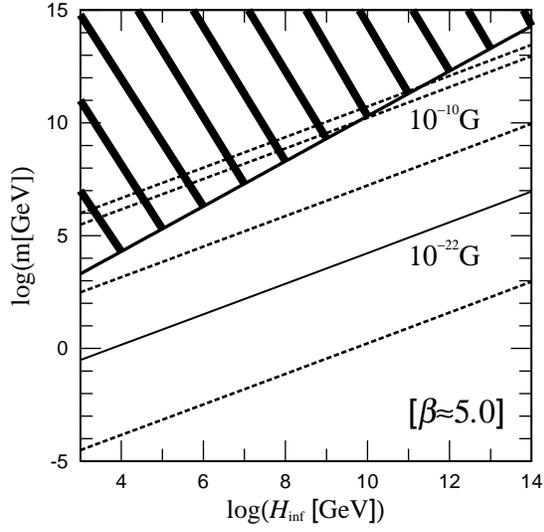}
\caption{The curves (dotted lines and a solid line) in the 
$H_\mathrm{inf}-m$ parameter space on which the present magnetic fields on 
1Mpc scale with each strength could be generated for the case 
$\beta \approx 5.0$. 
$\tilde{B}(t_0)=10^{-9} \mathrm{G}$, $10^{-10} \mathrm{G}$, 
$10^{-16} \mathrm{G}$, $10^{-22} \mathrm{G}$ (solid line), and 
$10^{-30} \mathrm{G}$ are shown (top down). 
The shaded area illustrates the region with 
$m > 2H_\mathrm{inf}$, where $t_\mathrm{R} > t_\mathrm{osc}$ 
and our analysis does not apply.  
Here we have taken the maximum of 
$ \tilde{\lambda} \kappa |{\Phi}_\mathrm{R}|$ for each case, and 
$w =0.01$ and ${\tilde{\lambda}} \sim \mathcal{O}(1)$.  
}
\end{center}
\end{figure}

\end{document}